\documentclass{aa}  

\usepackage{graphicx}
\usepackage{txfonts}
\newcommand{\flag}[1]{\texttt{#1}}
\usepackage{hyperref}
\hypersetup{colorlinks=true,linkcolor=blue,citecolor=blue,urlcolor=blue}
\usepackage{orcidlink}

\begin{document}


\title{Co-evolution of the Milky Way high- and low-$\alpha$ sequences
\\with chemical evolution models}
   \titlerunning{Co-evolution}

   \author{V. Grisoni \orcidlink{0000-0001-7366-7699}
          \inst{1,2}
          \and
          E. Spitoni  \orcidlink{0000-0001-9715-5727} \inst{1,2}
          \and
          F. Matteucci \orcidlink{0000-0001-7067-2302}  \inst{1,3,4}
          }

   \institute{INAF, Osservatorio Astronomico di Trieste, via G.B. Tiepolo 11, I-34131, Trieste, Italy\\
              \email{valeria.grisoni@inaf.it}
         \and
    IFPU, Institute for Fundamental Physics of the Universe, Via Beirut 2, 34151 Trieste, Italy 
            \and
           Dipartimento di Fisica, Sezione di Astronomia, Università di Trieste, Via G. B. Tiepolo 11, 34143 Trieste, Italy 
           \and
            INFN Sezione di Trieste, via Valerio 2, 34134 Trieste, Italy 
             }

   \date{Received --; accepted --}

 
  \abstract
   {Observational data have revealed a clear dichotomy in the [$\alpha$/Fe] vs. [Fe/H] diagram of the Milky Way thick and thin disc stars. Many recent studies have shown evidences of a co-evolution phase between the high- and low-$\alpha$ disc sequences as well as the presence of very old low-$\alpha$ stars.}
   {We aim to revise the parallel chemical evolution model that assumes two parallel histories of star formation for the two discs, by considering a pre-enriched delayed second infall episode in our revised scenario. By means of our chemical evolution models, we aim to explore the effects of a phase of co-evolution and the presence of old low-$\alpha$ stars, as recently observed.}
   {We consider a new version of the  parallel scenario for the  Milky Way thick and thin disc formation, which consists into two distinct infall episodes of slightly pre-enriched gas. The gas is considered to be  extragalactic but possibly contaminated by chemically enriched gas of a massive dwarf galaxy as Gaia-Enceladus, which merged with the Milky Way at least 10 Gyrs ago.
   Moreover, we test in our model observationally derived star formation histories of kinematically selected thick and thin discs, suggesting that the star formation is triggered by the passages of the Sagittarius galaxy.}
   {Our models can well explain the [$\alpha$/Fe] vs. [Fe/H] diagram from APOGEE DR17. The high-$\alpha$ sequence should have formed fast (less than one billion years), whereas the low-$\alpha$ one on a much longer timescale in the solar vicinity (several Gyrs). A hiatus arises as a period of low star formation between the formation of the high- and low-$\alpha$ sequences.  We are able to predict the existence of low-$\alpha$ stars older than 11 Gyrs, as found in the considered observational sample. Concerning the observationally motivated histories of star formation, we find that the star formation rate for the thin disc can well reproduce the data also with the inclusion of bursts; on the other hand, a prolonged star formation history for the thick disc is not compatible with its observed stellar age distribution of a very old population.} 
  {Our revised chemical evolution model with a pre-enriched and delayed  (roughly 1 Gyr) second infall episode, explains not only the abundance patterns of high- and low-$\alpha$ stars but also stellar age distributions for the selected observational sample. We predict a short co-evolution period in between the two phases and we can explain  the observed old low-$\alpha$ stars, but still further data for precise stellar ages would be needed to put more stringent constraints on their physical nature.}


   \maketitle
%

\section{Introduction}

Galactic archaeology aims to reconstruct the history of formation and evolution of the Galaxy (\citealt{Matteucci2021} and references therein).
It is now well-established that the Milky Way has a disc structure, with geometrically different thick and thin discs \citep{Gilmore1983}. It has been found that the Milky Way thick and thin discs not only differ in their kinematical and physical structure, but also in chemical abundances with a clear dichotomy in the [$\alpha$/Fe] vs. [Fe/H] relation, with thick-disc stars being generally $\alpha$-enhanced with respect to the thin-disc ones \citep{RecioBlanco2014,Hayden2015,Mikolaitis2017,Vincenzo2021}. Also, they differ on the basis of stellar ages, with thick-disc stars being generally older than thin-disc stars \citep{SilvaAguirre2018,Miglio2021}.
Thus, the Milky Way thick and thin discs can differ in structural, chemical, kinematic and age properties. Different selection methods exist to classify stars as thin or thick disc stars, each with its own advantages and limitations (see \citealt{Alinder2025} for an in-depth discussion on different definitions of thick and thin discs and their applications). In this work, we will focus on the chemical bimodality between the Milky Way thick and thin discs, and thus investigate the Milky Way discs as chemically defined and refer to them as high-$\alpha$ and low-$\alpha$ disc sequences, respectively.
\\In this context, chemical evolution models are powerful tools to understand how our Galaxy formed and evolved, and put strong constraints on the timescales of formation of the various Galactic components (\citealt{Matteucci2001,Matteucci2012,Kobayashi2020}).
In the literature, many Galactic chemical evolution models have been developed to follow the chemical evolution of the Milky Way and they have passed through different approaches: i) stochastic approach \citep{Argast2000,Cescutti2008,Rizzuti2025,Grisoni2025}; ii) serial approach \citep{Matteucci1986,Matteucci1989} iii) parallel approach \citep{Ferrini1992,Pardi1995,Chiappini2009,Grisoni2017}; iv) two-infall approach \citep{Chiappini1997,Romano2010,Spitoni2019,Spitoni2024,Dubay2025}; v) radial migration approach \citep{Schoenrich2009,Minchev2013,Kubryk2013,Kubryk2015,Johnson2021,Sharma2021,Chen2023,Prantzos2025}.
In the serial approach, it is assumed that the halo, thick and thin
disc form in sequence. In this case, the thick disc is just
a later phase relative to the halo, and the thin disc is a later phase
relative to the thick disc. The two-infall model belongs to the serial
approach, but it assumes that the halo and thick disc formed out of a
completely independent gas accretion episode relative to the thin
disc. Another sequential scenario similar to the two-infall model is the three-infall one, assuming a further infall episode forming the stellar halo (\citealt{Micali2013}) or occurring during the thin disc formation  (\citealt{Spitoni2023}).
An alternative way to model the evolution of different Galactic components is to follow their evolution not sequentially but in parallel (see \citealt{Ferrini1992,Pardi1995,Mishenina2006,Chiappini2009,Grisoni2017,Grisoni2018,Goswami2021}). In particular, \cite{Grisoni2017} proposed the "parallel model" specifically tailored to follow the evolution of the Galactic thick and thin discs, which has been shown to reproduce well the [Mg/Fe] vs. [Fe/H], and then also other abundance patterns, from the light elements up to the heaviest ones (\citealt{Grisoni2019,Grisoni2020a,Grisoni2020b,Grisoni2021}). In this model, it assumed that the Milky Way thick and thin discs are formed through two distinct infall episodes (namely, two separate one-infall models), that evolve at different rates, with different star formation histories (SFH), but they start forming stars at the same time. In this way, it is possible to capture a period of possible co-evolution for the two components.
\\In the last years, stellar ages have become another fundamental constraint for the chemical evolution of the Milky Way (\citealt{Soderblom2010} for a review) and different methods have been used, such as isochrone fitting \citep{Haywood2013,Hayden2017,Queiroz2023,Cerqui2025}, asteroseismology \citep{Anders2017b,Anders2017a,SilvaAguirre2018,Miglio2021,Willett2023,Willett2026,Warfield2024}, chemical clocks \citep{Casali2023,Casali2025,Roberts2025}. In order to explain the data of \cite{SilvaAguirre2018} for the stellar ages of Milky Way high- and low-$\alpha$ disc stars, \cite{Spitoni2019} applied the two-infall model and showed that there should be an important delay for the second infall episode ($\sim$4.3 Gyr, see also \citealt{Palla2020,Spitoni2021}), as opposed to the delay of $\sim$ 1 Gyr found by \cite{Chiappini1997}. However, recent studies showed that there could be a time overlap between the formation of these two components \citep{Beraldo2021,Wu2023,Gent2024}, thus suggesting that there could be a period of co-formation that cannot be explained by purely sequential chemical evolution models like the two-infall (see also \citealt{Dubay2025}). On the other hand, the co-evolution could be naturally interpreted in the framework of a parallel chemical evolution model.
Moreover, the presence of a group of old low-$\alpha$ disc stars with ages older than 9 Gyr has been found in several independent studies \citep{Haywood2013,Hayden2017,SilvaAguirre2018,Laporte2020a,Ciuca2021,Beraldo2021,Queiroz2023,Gent2024,Nepal2024,Casali2025,Borbolato2025}. This is also supported by the RR Lyrae stars possessing thin-disc orbits and chemistry  \citep{Prudil2020,Crestani2021,Orazi2025,Zhang2025,Bono2026}.  \\Among the first studies finding the presence of old low-$\alpha$ stars, \cite{Haywood2013} and \cite{Hayden2017} showed significant
temporal overlap of the high-$\alpha$ and low-$\alpha$ components of the Galactic disc. \cite{SilvaAguirre2018} also found a population of old low-$\alpha$ disc stars in the APOKASC sample; however, their age uncertainities were very large (with an uncertainity of $\sim$5 Gyr at an age of 12 Gyr) and no firm conclusions could be drawn on that population. Using APOGEE
and SEGUE data, \cite{Laporte2020a} also reported the existence of old
low-$\alpha$ stellar populations with ages between 8-12 Gyr in the
Anticenter Stream. Moreover, many other subsequent studies found old low-$\alpha$ disc stars \citep{Ciuca2021,Beraldo2021,Gent2024}.We also note that old low-$\alpha$ populations defined chemically (as already noted by \citealt{Emma2025}, \citealt{Arentsen2024}) can also be likely contaminated by accreted populations, whose low-energy tail is interpreted as old low-$\alpha$ (thin) disc \citep{Viswanathan2025}.  Thus, the presence of old low-$\alpha$ stars is currently matter of lively debate, and the question of how and when do stable and dominant galactic stellar discs form still remains a hot topic in the field (see e.g. \citealt{Zhang2024}).
\\Recently,
\cite{Nepal2024} using Gaia-RVS carried out a detailed chrono-chemo-dynamical study of a large sample of stars with precise stellar parameters, focusing on the oldest stars in order to decipher the assembly history of the Milky Way discs and found that the Galaxy, similar to the high-z galaxies observed by JWST and by ALMA, has an old thin disc. In particular, they confirmed the existence of metal-poor stars in thin-disc orbits: the majority of those stars are predominantly old
(>10 Gyr), with over 50\% being older than 13 Gyr.
More recently, \cite{Borbolato2025} (hereafter B25) concluded that there is an old chemically defined thin disc population with ages $>$11 Gyr that indicates a period of co-formation between the
high- and low-$\alpha$ sequences of the Milky Way disc. 
Finally, recently many efforts have been done in constraining the SFHs of the Galactic thick and thin discs \citep{Ruiz-Lara2020,Gallart2024,Emma2025}. In particular, \cite{Ruiz-Lara2020} observed kinematically selected thick and thin discs stars, and found a parallel evolution for these two Galactic components.
Recently \cite{Emma2025} presented new results for the star formation histories of the kinematically selected thick and thin discs of the Milky Way, showing also a short phase of co-evolution between the two components.
\\In this paper, we propose a new chemical evolution model based on the approach of \cite{Grisoni2017} to explain the recent observational data for the [$\alpha$/Fe] vs. [Fe/H] diagram as well as the age distributions of the high- and low-$\alpha$ disc populations of the Milky Way. In fact, \cite{Spitoni2019}, by analyzing APOKASC data, showed that a pure parallel model, where the two discs start forming exactly at the same time, predicts too many old low-$\alpha$ stars. Thus a revision of such model appears necessary to best reproduce all the latest observational constraints, including a co-evolution phase between the high- and low-$\alpha$ disc sequences and the presence of old low-$\alpha$ stars as found by B25. 
\\The paper is structured as follows. In Section 2, we outline the data to be compared with our chemical evolution models. In Section 3, we present the chemical evolution models used in this work. In Section 4, we show the results based on the comparison between data and model predictions. Finally, in Section 5 we summarize our conclusions.

\begin{figure*}
\centering
    \includegraphics[scale=0.3]{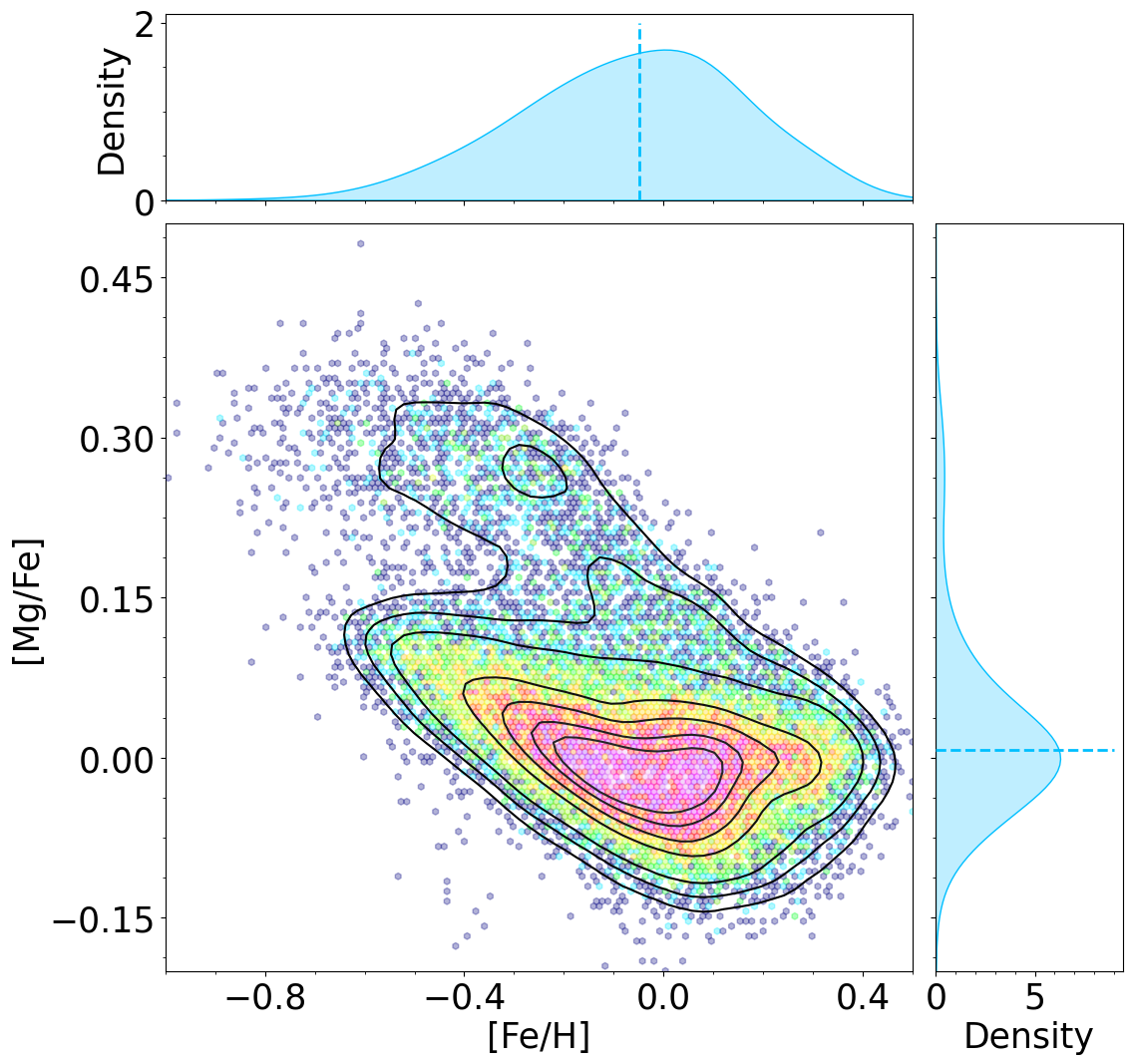}
    \includegraphics[scale=0.3]{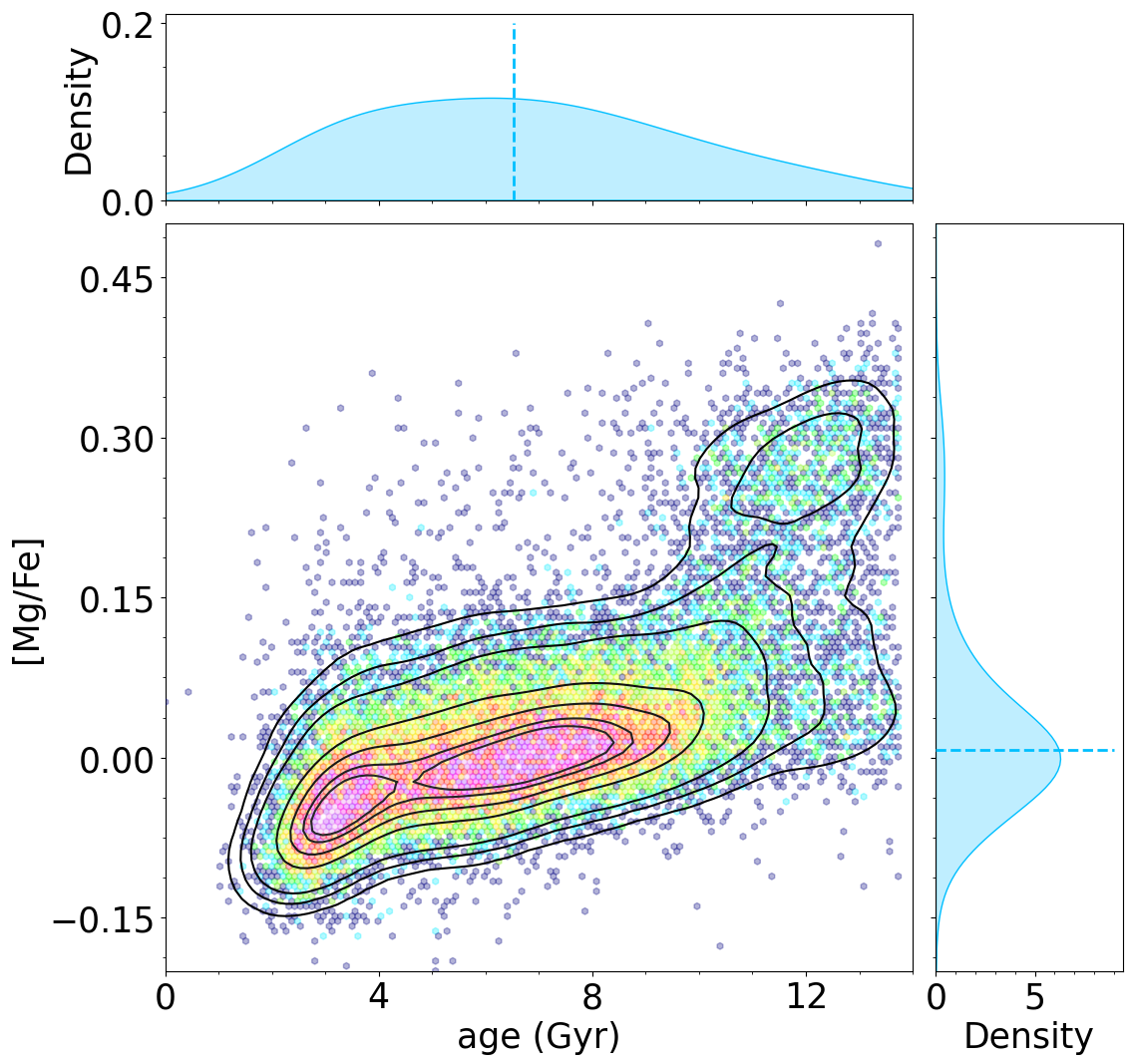}
    \caption{Observed [$\alpha$/Fe] vs. [Fe/H] (left) and [$\alpha$/Fe] vs. age (right) diagrams for the final sample from \cite{Borbolato2025} with chemical abundances from APOGEE DR17 and stellar ages from \flag{StarHorse}. The colour-coding is presented on a logarithmic scale. The contour lines enclose fractions of 
    0.95, 0.88, 0.8, 0.60, 0.45, 0.30, 0.20 of the total number of observed stars. In the marginal panels, we report the corresponding distributions with their medians. Details on the data selection are specified in the text.}
    \label{APOGEE}
\end{figure*}

\section{Observational data}


In this Section, we describe the observational data from B25 which have been used in this work, both for chemical abundances and stellar ages to compare with the predictions of our chemical evolution models. We address the reader to B25 for further details on the observational sample.

\subsection{Spectroscopic data}
 We consider the APOGEE sample investigated in B25.
Spectroscopic data are thus from the high-resolution spectroscopic survey APOGEE \citep{Majewski2017}, in particular its latest data release APOGEE DR17 \citep{APOGEEDR17}.
APOGEE is part of the Sloan Digital Sky Surveys (SDSS) and it operates using the du Pont Telescope and the Sloan Foundation 2.5 m Telescope \citep{Gunn2006} at Apache Point
Observatory. Stellar parameters and chemical abundances are derived with the APOGEE Stellar Parameters and Chemical Abundance Pipeline (ASPCAP; \citealt{garciaperez2016}). Model atmospheres used in APOGEE DR17 are based on the MARCS model
(\citealt{gustafsson2008}, as discussed by \citealt{jonsson2020}),
and the line list is described in \cite{smith2021}.
In order to obtain stars with high-quality abundance measurements, B25 applied quality cuts, excluding sources that exhibited issues in the spectra, the spectral fitting process and the estimated parameters (\flag{STARFLAG} == 0, see \citealt{jonsson2020}). Moreover, they ensured accurate estimates of
the [Fe/H] and [Mg/Fe] abundances by selecting sources with no flagged
issues (namely, flagged == 0). They also selected stars with a good
signal-to-noise ratio (S/N > 50), which allows precise
estimates of chemical abundances, radial velocities, and stellar
parameters \citep{garciaperez2016,jonsson2020}.
\subsection{Stellar ages}
Stellar ages in B25 were derived by \cite{Queiroz2023} using the \flag{StarHorse} code, which is a Bayesian isochrone-fitting code. 
In particular, in \cite{Queiroz2023} they used \flag{StarHorse} for the first time to derive stellar ages for main-sequence turnoff (MSTO) and subgiant branch (SGB) stars, with age uncertainties typically around 30$\%$; the uncertainties drop to 15$\%$ for SGB only stars. Age estimates are reliable for MSTO and SGB stars, since isochrones of different ages are better separated in this region of the Hertzsprung-Russell diagram; however, the low luminosity of both MSTO and SGB stars leads to an age sample restricted to the solar neighborhood (d<2 kpc), which is good to constrain our chemical evolution models in the solar vicinity. In \cite{Queiroz2023}, they presented results for stars listed in both Gaia DR3 and various public spectroscopic surveys; as discussed in the previous subsection, we will focus on the APOGEE DR17 case, which has the smallest nominal uncertainties. B25 restricted their final sample to stars with age uncertainty $\sigma_{\text{AGE}}$ < 1 Gyr, even if they comment that the qualitative results are mantained if the age uncertainity cut is removed. 
\subsection{Final sample}
After all the quality cuts both for chemical abundances and stellar ages, the final sample contains 25,439
stars with APOGEE DR17 chemical abundances and derived ages from \flag{StarHorse} (we address the interested reader to the B25 paper for further details on the observational sample).
\\In the left panel of Fig. \ref{APOGEE}, we show the [$\alpha$/Fe] versus [Fe/H] for the B25 sample with APOGEE DR17 chemical abundances. We can clearly see the two distinct sequences corresponding to the thick and thin discs (high- and low-$\alpha$ respectively). We also show the [Fe/H] and [Mg/Fe] distributions; in particular, it is evident the bimodality in the [Mg/Fe] distributions corresponding to the high- and low-$\alpha$ sequences, with two distinct peaks in [Mg/Fe] distributions. We note the ratio of $\sim$ 1:10 between the number of high-$\alpha$ and low-$\alpha$ stars in the sample of B25 (see also \citealt{Emma2025} for a similar ratio between their kinematically selected thick and thin discs); this is different from e.g. \cite{Snaith2015} where the high-$\alpha$ consistutes $\sim$ 40\% of the disc population and the bimodality is more evident. To properly compare with the observational sample of B25, we will consider a similar ratio in our models (see next Section).
\\In the right panel of Fig. \ref{APOGEE}, we show the observed [$\alpha$/Fe] vs. age (Gyr) diagram in the solar vicinity for the B25 sample with APOGEE DR17 chemical abundances and stellar ages from \flag{StarHorse}. The high-$\alpha$ sequence is peaked towards old ages, being a very old component (see also \citealt{Miglio2021}), whereas the low-$\alpha$ sequence evolve much more slowly. We can note the presence of old low-$\alpha$ stars (ages older than 11 Gyr). However, a note of caution is warranted regarding the stellar ages from \cite{Queiroz2023} used in B25. Their age estimates rely on priors that explicitly allow for very old low-$\alpha$ stars, and thus the presence of very old low-$\alpha$ stars may partially reflect modelling assumptions and/or age uncertainty/systematics rather than independent evidence. However, in B25 they concluded that the presence of an ancient low-$\alpha$ population is not an artifact of the prior (see their Appendix A) and thus their conclusions on the low-$\alpha$ population do not change. Still, further data for precise stellar ages in particular for the oldest stars should be needed to put more stringent constraints on the earliest phases of the Galactic evolution.
\\We note that our analysis is then based on the comparison with the selected B25 sample and not to a fully selection-function-corrected intrinsic Milky Way population. In the following, we will discuss how the aforementioned dataset compare with our chemical evolution models for the thick and thin discs.


\section{Chemical evolution models}

\begin{table*}
	\centering
	\caption{Input parameters for the Galactic chemical evolution models used in this work for the Milky Way thick and thin discs.}
	\label{Tab1}
 	\begin{tabular}{lcccccr} 
		\hline
		Model	  & $\nu$ & $\tau$   & t$_{start}$&$(X_i)_{inf}$ \\
           & [Gyr$^{-1}]$ & [Gyr] &     [Gyr ago]& Mg, Fe \\
		\hline
Classical-Thick & 2 & 0.1 &  13.7 & -\\
Classical-Thin & 1 & 7 &  13.7 & - \\
Revised-Thick & 5 & 0.1  &  12.5 & 2.19$\cdot10^{-4}$, 2.32$\cdot10^{-4}$\\
Revised-Thin & 0.5 & 7     &  11.4 & 1.87$\cdot10^{-4}$, 3.22$\cdot10^{-4}$\\
		\hline
	\end{tabular}
    \tablefoot{In the first column, there is the name of each model. In the following columns, we list: the star formation efficiency ($\nu$ in Gyr$^{-1}$), infall timescale ($\tau$ in Gyr), the onset of formation (t$_{start}$ in Gyr ago) having that $\Delta_{{\rm T}}=1.2$ Gyr and  $\Delta_{t}=1.1$ Gyr for the thick and thin discs, respectively (see Eqs. (\ref{eq_1IMthick}) and (\ref{eq_1IMthin}) in Section 3 for details). In the last column  the presence of primordial or pre-enriched gas infall is indicated.}
 \end{table*}

\begin{figure*}
\centering
   \textbf{Classical parallel vs. revised}\par\medskip
 	\includegraphics[scale=0.5]{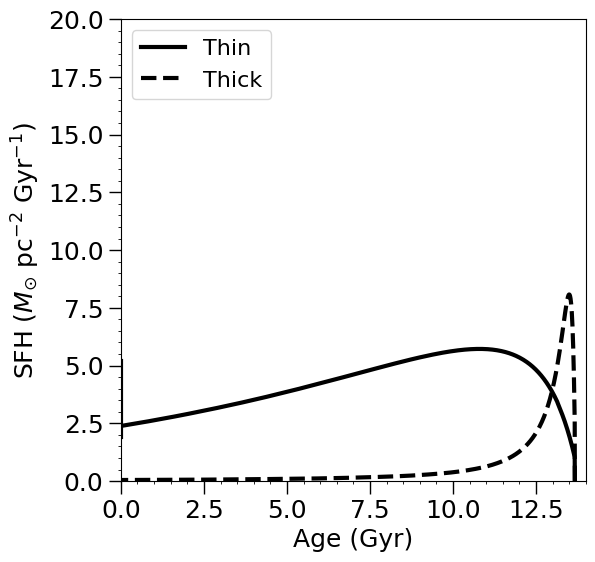}
    \includegraphics[scale=0.5]{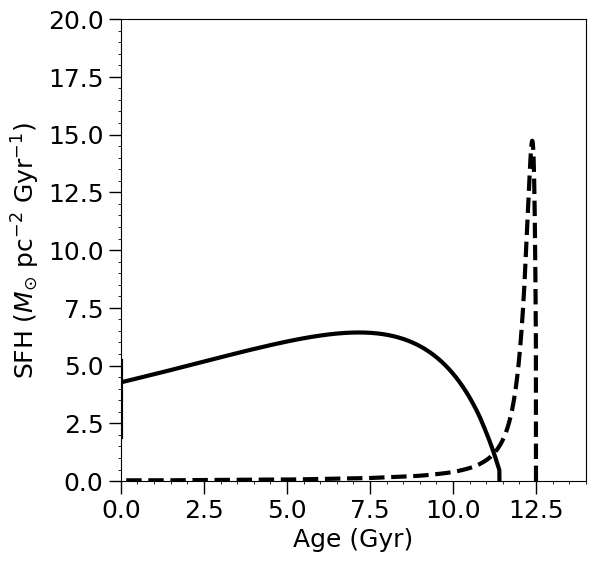} 
    \caption{Star formation histories (SFHs) for the chemical evolution models of the thick and thin discs considered in this work (the classical parallel on the left panel and the revised model on the right). The black bar is the present day star formation rate observed in the thin disc.
    }
    \label{SFH}
\end{figure*}


To model the chemical evolution of the Milky Way thick and thin discs, we start from the parallel model of
\cite{Grisoni2017}. The model has been extended to all the Galactic disc in
\cite{Grisoni2018} and tested on the abundance patterns of several chemical elements, from the light ones (H, He, D, and Li) to the heavy ones  ($\alpha$-elements, Fe-peak-elements, s- and r- process elements, see e.g. \citealt{Grisoni2019,Grisoni2020a,Grisoni2020b,Grisoni2021,Grisoni2024_proceeding}). Here, we will constrain our models also in the light of data from stellar ages.
\\In the classical parallel approach proposed by \citet{Grisoni2017}, it is assumed that the thick and  thin disc of the Galaxy form via two distinct infall events starting at the same time and evolving at different rates. 
In this article, we introduce a more general formulation of the gas infall terms that allows for different durations of the disc phases, which can be written as:


\begin{equation} \label{eq_1IMthick}
(\dot G_i(R_{\odot},t)_{inf})|_{thick}=\theta(t-{\Delta}_{{\rm T}}) \, A(R_{\odot}) \, (X_i)_{inf} \, e^{-\frac{t-\Delta_{{\rm T}}}{\tau_1}},
\end{equation}
and
\begin{equation} \label{eq_1IMthin}
(\dot G_i(R_{\odot},t)_{inf})|_{thin}=\theta(t-{\Delta}_{{\rm T}} -\Delta_{ t}) \, B(R_{\odot})\,(X_i)_{inf} \, e^{-\frac{t-\Delta_{{\rm T}}-\Delta_{t} }{\tau_2}},
\end{equation}
where  $\theta$ is  Heaviside step function.  
Defining  $t_G=13.7$ Gyr  as the evolutionary time of the whole Galaxy  and $t_{{\rm T}}$ as that of the thick-disc phase, we have that   ${\Delta}_{{\rm T}}=t_G-t_{{\rm T}}$, i.e. the delay between the onset of the Galaxy evolution and the start of the thick-disc evolution. In eq. (\ref{eq_1IMthin}), the quantity  $\Delta_{t}$ is the delay in the thin disc onset compared to the one of the thick disc.  
We recover the classical parallel scenario, in which the two discs begin forming simultaneously at the onset of Galactic evolution, by imposing  $\Delta_{{\rm T}}=\Delta_{t}  =0$. 
\\We adopt $\tau_1$ equal to 0.1 Gyr for the thick disc, and $\tau_{2}$ is equal to 7 Gyr for the thin disc at the solar neighbourhood \citep{Grisoni2017}. 
The parameters $A(R_{\odot})$ and $B(R_{\odot})$ are set to reproduce the present-time total surface mass densities in the solar vicinity for the two discs, as suggested by \cite{nesti2013}; in particular, this is equal to 65 M$_{\odot}$pc$^{-2}$ for the thin disc and 6.5 M$_{\odot}$pc$^{-2}$ for the thick disc. As discussed in the previous Section, we note the assumed ratio 1:10 between the thick and thin discs components (as already considered in \citealt{Grisoni2017}), which also reflects the ratio in the observed numbers of high- and low-$\alpha$ stars from the sample of B25. However, we recall that different ratios can be present in the literature (see e.g. \citealt{Spitoni2021} where the ratio between $A(R_{\odot})$ and $B(R_{\odot})$ has been considered a free parameter of the model and different values were tested there). Here, for consistency with the observational sample that we want to explain, we fix 65 M$_{\odot}$pc$^{-2}$ for the present-time total surface mass density of the thin disc and 6.5 M$_{\odot}$pc$^{-2}$ for the thick disc, as done also in \cite{Grisoni2017}.
$(X_i)_{inf}$ is the composition of the infalling gas. We start by assuming primordial gas infall, as originally done in the classical parallel of \cite{Grisoni2017}, but we test also the effect of pre-enriched gas with contamination by the chemically enriched gas of a massive dwarf galaxy which merged with the Milky Way, following the approach described in \cite{Spitoni2024}, see Appendix for details.
\\For the initial mass function (IMF), we take the one of \cite{Kroupa1993}, as in \cite{Grisoni2017}. 
The star formation rate (SFR) follows the Schmidt-Kennicutt law \citep{Kennicutt1998}:
\begin{equation} \label{eq_03_02}
\psi(t)=\nu_{1,2} \, \sigma_{gas}^k,
\end{equation}
where $\sigma_{gas}$ is the surface gas density, $k$=1.5 is the law index and $\nu_{1,2}$ is the star formation efficiency for the disc phases. Following \cite{Grisoni2017}, we assume higher star formation efficiency in the thick disc with respect to the thin disc.
We also test the effect of possible bursts of star formation, as suggested by the star formation history of \cite{Ruiz-Lara2020} where three main bursts of star formation are found in the thin disc and are associated with possible passages of Sagittarius dwarf galaxy. We model these bursts as periods of enhanced star formation efficiency (see also \citealt{Prantzos2025}).
\\We note that in this paper we focus on the chemical evolution in the solar neighborhood and thus run our chemical evolution models in the solar vicinity (at R= R$_{\odot}$=8 Kpc); however, the parallel scenario can be also extended to other Galactocentric distances by including the inside-out scenario, as done in \cite{Grisoni2018}. Stellar radial migration has also been tested in the context of Galactic chemical evolution models in \cite{Palla2022}, but they showed that it has a small effect on the overall distribution functions in the solar vicinity (see also \citealt{Vincenzo2021,Khoperskov2021,Spitoni2025}).


\subsection{Nucleosynthesis prescriptions}

In this work, we adopt nucleosynthesis prescriptions following \cite{Grisoni2017}, which are based on model 15 of \cite{Romano2010} where an exhaustive description of the adopted yields can be found.
In particular, the nucleosynthesis prescriptions are as follows. For single stars with mass 0.8–8 M$_{\odot}$, we adopt the nucleosynthesis prescriptions of \cite{Karakas2010}.
For SNe Ia, we take into account the nucleosynthesis prescriptions of \cite{Iwamoto1999}.
For massive stars with masses M $>$ 8 M $_{\odot}$, we adopt the nucleosynthesis prescriptions of \cite{Kobayashi2006} for the following elements: Na, Mg, Al, Si, S, Ca, Sc, Ti, Cr, Mn, Co,
Ni, Fe, Cu and Zn. As for the He and CNO elements, we consider the results of Geneva models for rotating massive stars (see \citealt{Romano2010}).
However, for Mg that is the most relevant element in this study, in order to
obtain a better agreement with the data we adopt yields of massive stars multiplied by a factor 1.2. Also as suggested in \cite{Matteucci2019}, we consider a multiplying factor for SNIa by a factor 10. It is well known, in fact,
that Mg yields have been underestimated in many nucleosynthesis
studies (see \citealt{Francois2004} for a discussion of this point), and
although the most recent ones have improved, the Mg production
is still underestimated and needs correction factors.

\begin{figure*}
\centering
   \textbf{Classical parallel vs. revised}\par\medskip
   \includegraphics[scale=0.3]{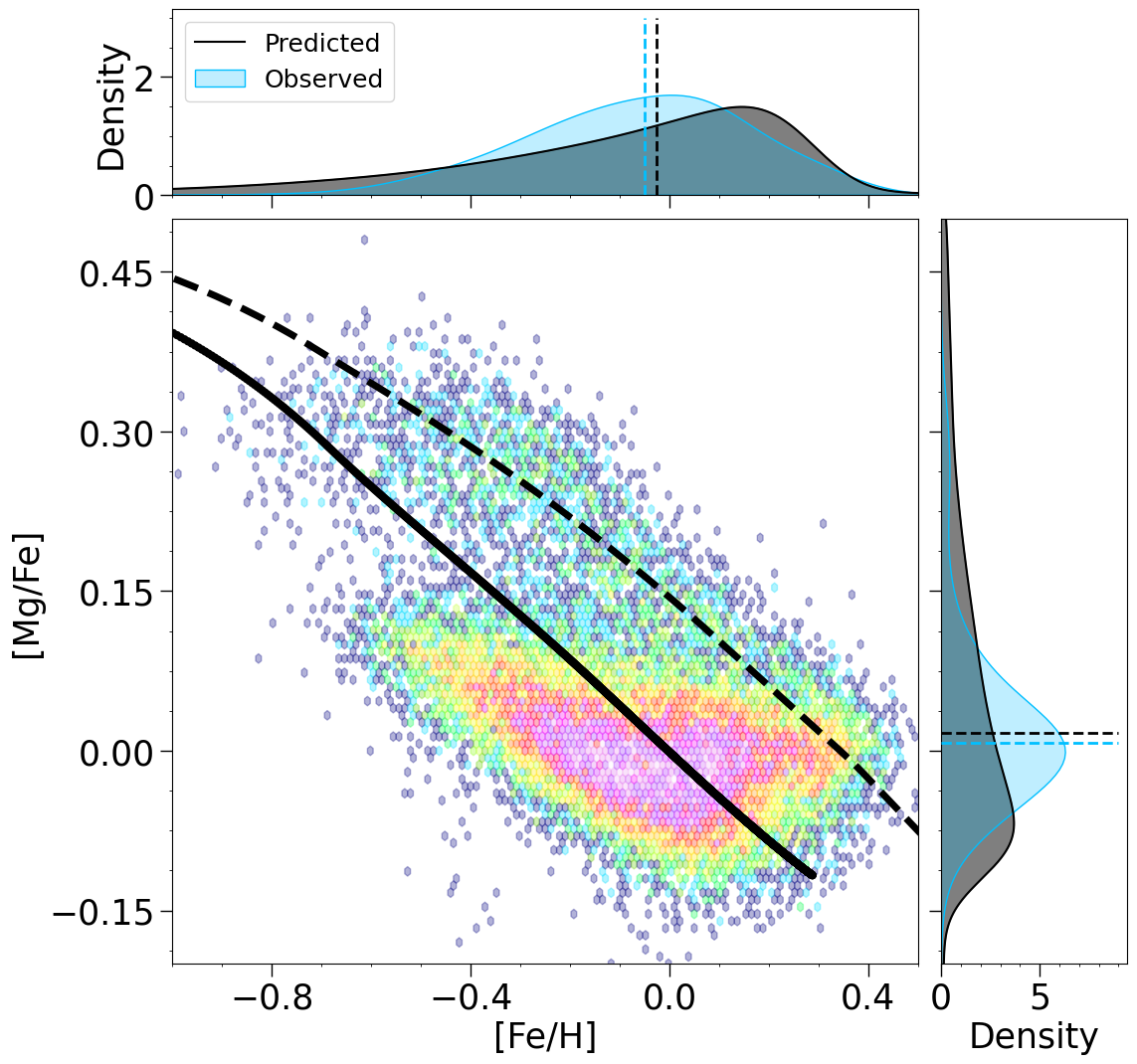}
 	\includegraphics[scale=0.3]{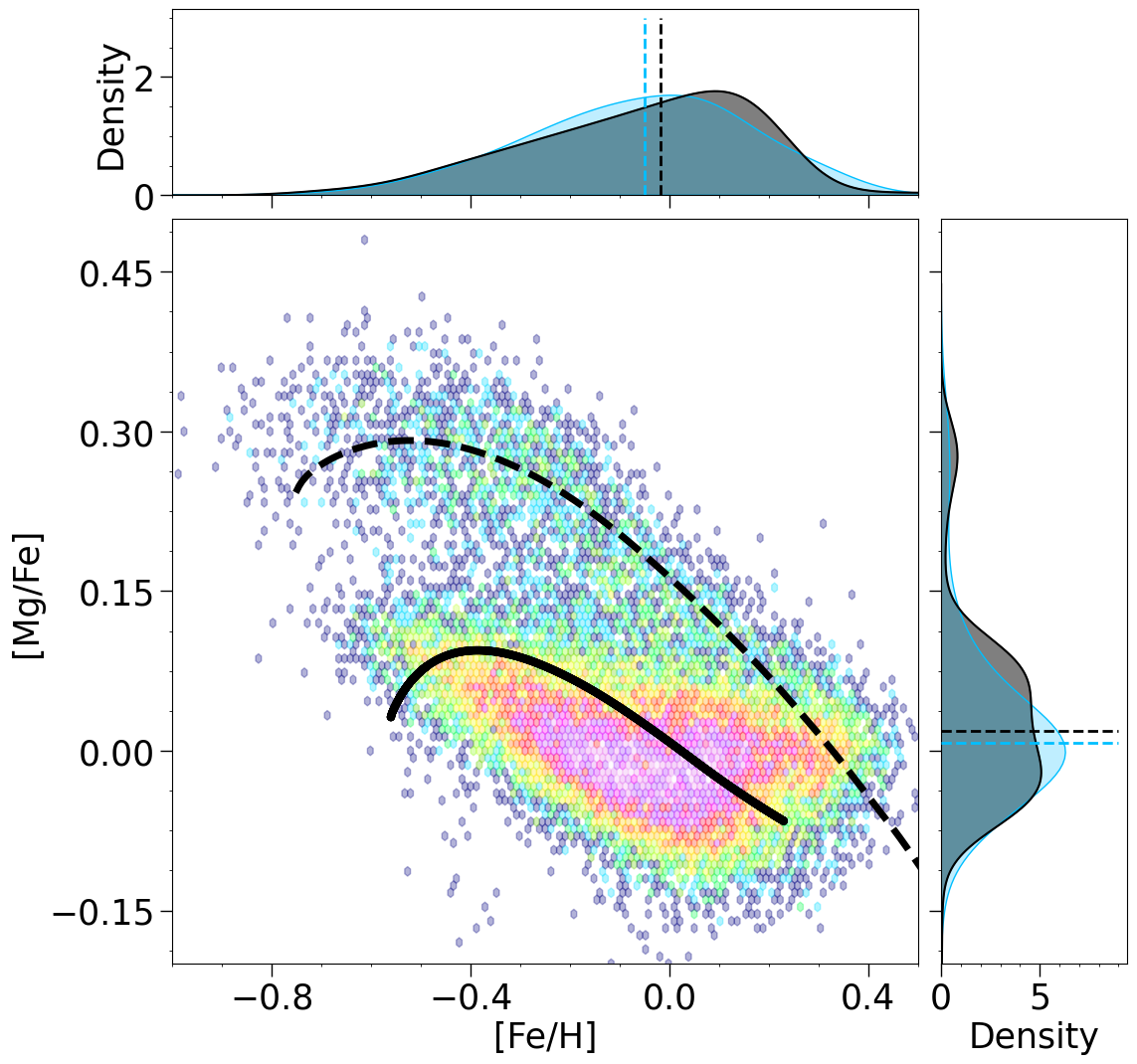}
   \caption{Observed and predicted [$\alpha$/Fe] vs. [Fe/H] diagram in the solar vicinity. Data are from the final sample of \cite{Borbolato2025} with chemical abundances from APOGEE DR17. The predictions are from our chemical evolution models, both the classical parallel model (left panel) and the revised one (right panel) for the Galactic thick disc (dashed black line) and thin disc (continuous line). On the sides of each panel, the observed (light blue shaded area) and predicted (dark grey shaded area) normalised KDEs of the distributions calculated with a Gaussian kernel are reported, together with their corresponding medians.}
    \label{MgFe_FeH}
\end{figure*}

\section{Results}

Here, we present our results, based on the comparison between data and model predictions in the framework of a parallel chemical evolution scenario for the Milky Way thick and thin discs. In Table 1, we show the models and their parameters, both the classical parallel and the revised one. In particular, in column 2, we report the star formation efficiency $\nu$ for the thick and thin disc in each model. In column 3, we show the timescale of gas infall for the thick and thin discs  and, in column 4, the time at which  the two discs started forming. Finally, in column 5, we indicate if the infalling gas is assumed to be primordial or pre-enriched (see details in Section 3). In the following, we discuss the results of of both the classical and the revised parallel model.
\subsection{Star formation histories}
In Fig. \ref{SFH}, we show the different star formation histories (SFH), as predicted by the two models, both the classical and revised parallel one, which are listed in Table 1.
In the case of the classical parallel of \cite{Grisoni2017}, we have two parallel events which start forming at the same time (same t$_{start}$), but they evolve at different rates (namely, with different star formation efficiencies and different infall timescales, faster in the thick disc than in the thin disc). The SFR of the thin disc is constrained by the observed present-day value, whereas the SFR of the thick disc goes rapidly to zero, and there is negligible star formation in the thick disc at present time. However, such an early onset t$_{start}$ for both the disc components was found to be problematic to reproduce age distributions for the thick and thin discs, as discussed in \cite{Spitoni2019} in comparison to APOKASC data by \cite{SilvaAguirre2018}. A prolonged co-evolution would be also problematic in the light of dynamical models where two gas streams feeding the disc at the same radius would then mix on short timescales. For these reasons, we thus revise the parallel scenario suggested by \cite{Grisoni2017}. We then show the predicted SFH of our revised parallel model (right panel), with a shortly delayed onset for the second infall episode (roughly 1 Gyr, as in the original two-infall formulation by \citealt{Chiappini1997}). Also in this case, the SFR of the thin disc is constrained by the observed present-day value, and the SFR of the thick disc decreases even more rapidly, with negligible star formation in the thick disc at present time. In the case of the revised model, the formation of the thick disc is even faster than the classical model (we assume higher star formation efficiency $\nu=$5 Gyr$^{-1}$), and we have a very old population, as  suggested by several studies \citep{Miglio2021,Queiroz2023}. On the other hand, the thin disc evolve more slowly, with lower star formation efficiency and longer infall timescale with respect to the thick disc. In the case of the revised parallel model, there is a longer period of low star formation between the formation of the thick disc and the onset of the thin disc, which is delayed of roughly 1 Gyr similarly to the original two-infall formulation by \cite{Chiappini1997}. This is a more reasonable scenario in the light of dynamical simulations, where two discs starting forming exactly at the same time and evolving in parallel as in the classical parallel formulation could be problematic to explain. Compared to the classical parallel model, this short delay produces a longer hiatus between the peaks of the SFH of the thick and thin discs,  a scenario similar to that discussed by \cite{Haywood2016}.
\\We see that in both models (the classical parallel and the revised one), the two disc components have a period of co-evolution as suggested by several studies \citep{Beraldo2021,Wu2023,Gent2024}, even if much shorter in the revised parallel case.
In each case, the thick disc forms faster (less than a billion years), whereas the thin disc forms on a much longer timescale of formation in the solar vicinity (7 Gyrs) and inside-out. 

\subsection{The [$\alpha$/Fe] vs. [Fe/H] diagram}

\begin{figure*}
\centering
   \textbf{Classical parallel vs. revised}\par\medskip
 	\includegraphics[scale=0.3]{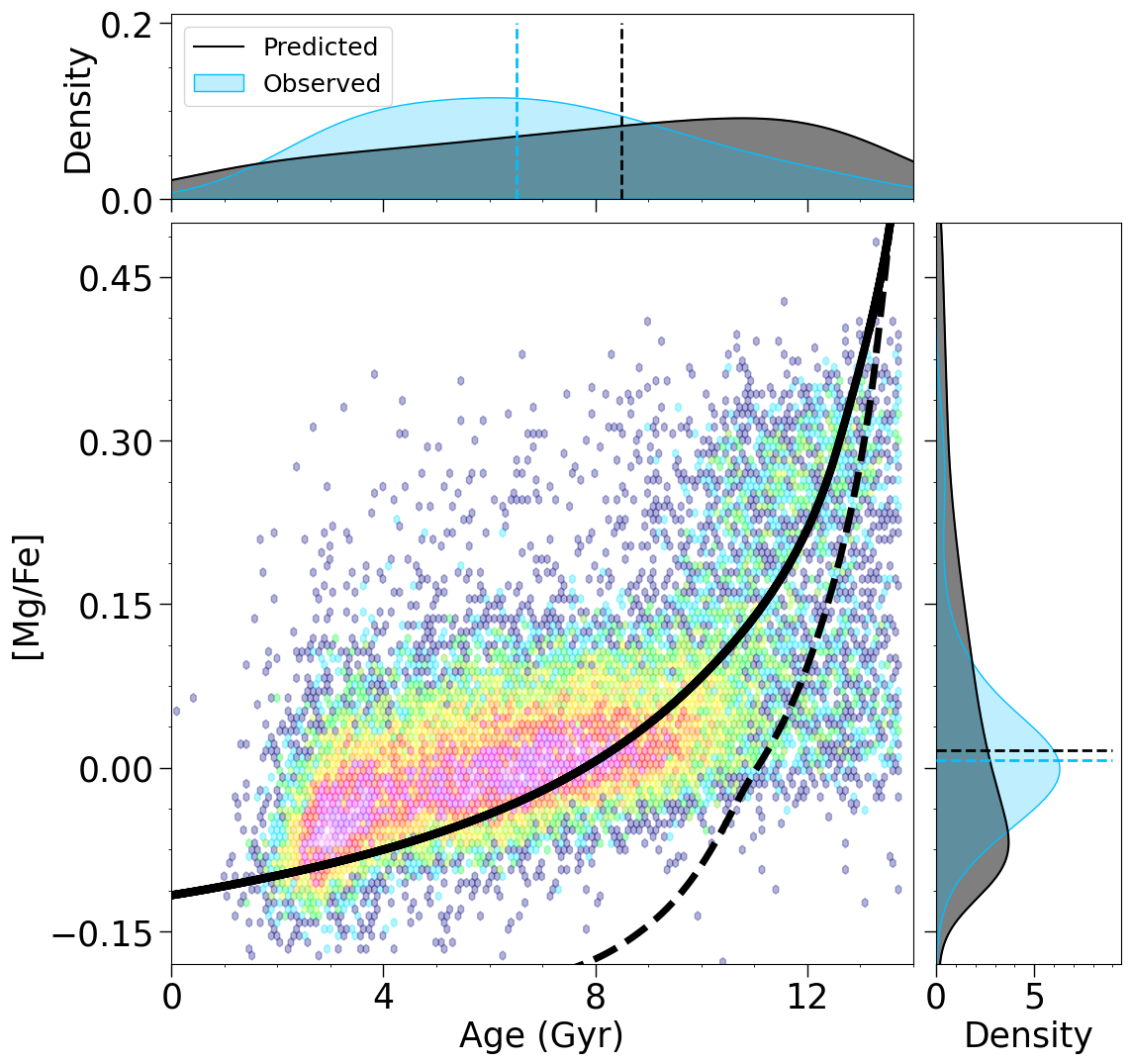}
                \includegraphics[scale=0.3]{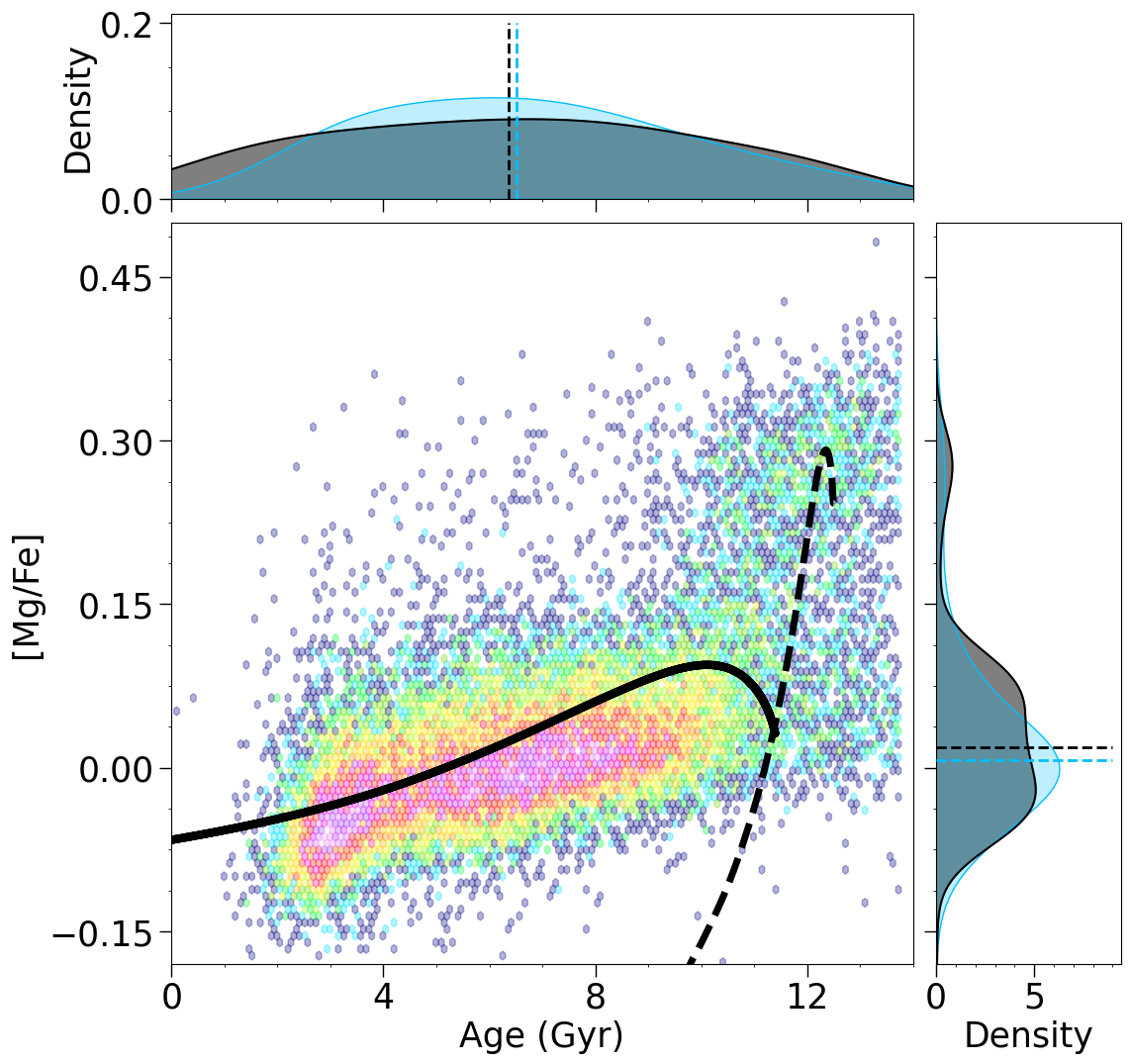}  
    \caption{Observed and predicted [$\alpha$/Fe] vs. age. Data are from the final sample of \cite{Borbolato2025} with chemical abundances from APOGEE DR17 and stellar ages from \flag{StarHorse}. The predictions are from our chemical evolution models, both the classical parallel model (left panel) and the revised one (right panel) for the Galactic thick disc (dashed black line) and thin disc (continuous line). On the sides of each panel, the observed (light blue shaded area) and predicted (dark grey shaded area) normalised KDEs of the distributions calculated with a Gaussian kernel are reported, together with their corresponding medians.
    }
    \label{MgFe_age}
\end{figure*}

In Fig. \ref{MgFe_FeH}, we show the [$\alpha$/Fe] vs. [Fe/H] diagram, both for the classical parallel (left panel) and the revised one (right panel).
In this plot with data from APOGEE DR17, it is evident the clear dichotomy between the thick and thin disc stars (low- and high-$\alpha$ sequences respectively).
We can see that the two high- and low-$\alpha$ sequences from APOGEE data can be well explained in the framework of the classical parallel model for thick and thin discs \citep{Grisoni2017}, with the high-$\alpha$ sequence forming on  a shorter timescale and higher star formation efficiency with respect to the low-$\alpha$ one. In fact, in both cases (both classical on the left and revised on the right), we can nicely see two distinct sequences for the thick and thin discs, with the thick disc being $\alpha$-enhanced due to its higher star formation efficiency and faster formation. However, when looking at the stellar density distribution as function of the [Mg/Fe] ratio (side panels of each figure), the two expected peaks in the [Mg/Fe] ratio are not clearly visible in the case of the classical parallel, where there was a long overlapping between the two components. On the other hand, a clear bimodality in the [Mg/Fe] distribution can be obtained in the case of the revised parallel model, where a short delay and a pre-enrichment in the gas are considered.
\\In the case of the revised parallel, following \cite{Spitoni2024}, we consider pre-enrichment levels for the thick  and thin discs. This allows us to best reproduce also the stellar metallicity ([Fe/H]) distribution function (MDF) of these Galactic components as well as the distributions as functions of [Mg/Fe].
To obtain such pre-enrichment levels, following \cite{Spitoni2024}, we modeled the evolution of a massive dwarf galaxy that can contribute to some pre-enrichment level (details of the chemical evolution model for a massive dwarf galaxy are discussed in more detail the Appendix A).  Such a pre-enrichment is particularly important for reproducing the initial low [Mg/Fe] ratios in metal-poor thin disc stars.
For the thick disc, following suggestions of \cite{Spitoni2024} to better reproduce also the MDF of thick disc stars, we consider a level of pre-enrichment for [Mg/Fe] and [Fe/H] reached after 1.2 Gyr of evolution of the massive dwarf galaxy model. This produces a bend in the [$\alpha$/Fe]–[Fe/H] track for the high-$\alpha$ sequence and allows us to better reproduce its [Mg/Fe] distribution (see also Fig. 7 in \citealt{Spitoni2024} for the impact of the pre-enrichment in the onset of the thick disc phase). For the thin disc, we consider a level of pre-enrichment reached after 2.3 Gyr of evolution of the massive dwarf galaxy model, but with a dilution factor of 50$\%$ due to the contemporary presence of primordial gas infall. In fact, we warn that the thin disc is mainly formed by extragalactic gas infall contaminated by the metal enriched gas of the dwarf galaxy. This scenario is supported by simulations, such as VINTERGATAN \citep{Renaud2021,Agertz2021} where the low-$\alpha$ sequence in a Milky Way-like galaxy is fuelled by the accretion of metal-poor gas from the circumgalactic medium (CGM). The accretion of this metal-poor gas from the CGM is seen as the cause of the metallicity overlap between the high-$\alpha$ and low-$\alpha$ stellar sequences observed in the Solar neighborhood of the Milky Way. This process is often associated with a second, delayed infall of gas in galaxy formation models, which dilutes the previously enriched interstellar medium (ISM) and fuels the formation of the low-$\alpha$ population. Our scenario do not require strong dilution levels as in the two-infall scenario, which have been shown to be problematic  \citep{Orkney2025,Johnson2025}. The final $(X_i)_{inf}$ values for Mg and Fe considered as pre-enrichment level in our chemical evolution models are reported in Table 1. Different levels of pre-enrichment reflect into the fact that the two model curves for the revised parallel case start at different [Mg/Fe] and [Fe/H] values. In summary, we can see that with our revised chemical evolution model we can nicely reproduce not only the two sequences in [$\alpha$/Fe] vs. [Fe/H], but also the two peaks in the [Mg/Fe] distribution, and thus well explain the $\alpha$-bimodality (see Appendix B for a separate analysis for the revised parallel model for thick and thin discs components).

\subsection{Comparison with stellar ages}
We now discuss in detail the predictions for the two models compared not only to abundance patterns, but also to age distributions from the observational sample of B25, showing a phase of co-evolution and the presence of old low-$\alpha$ stars that can be explained by a parallel chemical evolution model. However, as discussed in the Introduction, the need for a revised parallel model resides in the fact that it predicts too many very old low-$\alpha$ stars. In fact, \cite{Spitoni2019} had found that the classical parallel model of \cite{Grisoni2017} with two discs starting forming exactly at the same time could not reproduce the observed stellar age distributions from APOKASC data of \cite{SilvaAguirre2018}.
\\In Fig. \ref{MgFe_age}, we show the [$\alpha$/Fe] vs. age for the sample of B25 compared to the predictions of our chemical evolution models. In the left panel, we compare with the predictions of our classical parallel model of \cite{Grisoni2017}, with the two discs starting forming exactly at the same time and evolving in parallel. However, we see the problem raised by \cite{Spitoni2019} for the classical parallel model, with a  predicted bulk of old stars in the predicted age distribution, that are not present in the observed age distribution. Also, the model curve of the thin discs (continuous black line) starts at too high [Mg/Fe] and thus cannot explain low-$\alpha$ stars at old ages, and thus a revision is needed.
\\In the right panel of Fig. \ref{MgFe_age}, we then show the predictions of our revised parallel model. We can see now that with such a revised model we can match the observed age distributions much better than the classical parallel. In fact, with a short delay for the second infall episode, there is no more the bulk of old low-$\alpha$ stars found in the age distributions of the classical parallel.
Moreover, we note that with the assumed pre-enrichment levels, the low-$\alpha$ sequence starts in the low-$\alpha$ region as needed and not at too high [Mg/Fe] levels, typical of the early evolutionary phases when the pollution from Type Ia SNe is not yet present. In this way, this revised parallel scenario including a delayed second infall episode and pre-enrichment levels can explain the observational data, not only the observed abundance patterns but also age distributions. 
\\The very fast formation in the chemical evolution model of the thick disc implies that the high-$\alpha$ sequence is formed by very old stars, as confirmed by several studies \citep{Miglio2021,Queiroz2023}. On the other hand, in our chemical evolution model, the thin disc forms much more slowly. In the revised model, by considering an appropriate level of pre-enrichment, as mentioned above, the low-$\alpha$ sequence now starts at the required level of $\alpha$-enhancement. By assuming a short delay for the second infall episode, we avoid to have too many  old low-$\alpha$ stars, as also noted in \cite{Spitoni2019}, and we can better explain the observed age distributions.
In fact, at variance with the classical parallel model (where the two discs were starting forming exactly at the same time more than 13 Gyr ago) there is no more the bulk of old thin disc stars at very old ages, since the onset of formation for the thin disc is assumed to be slightly delayed with respect to the classical parallel (now the onset of the thin disc is assumed to be $\sim$11 Gyr ago). We thus note that our chemical evolution model for the thin disc can explain also very old stars forming early on in the low-$\alpha$ sequence as found in B25, that could not be explained in terms of a purely sequential scenario such as the two-infall model \citep{Spitoni2019,Spitoni2024}. 
\\It is worth noting that \cite{Spitoni2019} concluded that, given the very large uncertainities of the APOKASC catalogue by \cite{SilvaAguirre2018} especially for old ages, also the two-infall model taking into account the observational errors could reach some old low-$\alpha$ stars.
In particular, it was shown that, once the observational uncertainties from the APOKASC catalogue were taken into account (see Figure 6 in \citealt{Spitoni2019}, where at an age of 12 Gyr there is an uncertainity of $\sim$5 Gyr), the predicted low-$\alpha$ phase could extend to ages as old as $\sim$12 Gyr. However, if the observational uncertainities are lower (of the order of $\sim 20\%$), a sequential two-infall scenario with a large delay for the second infall episode starting  $\sim$9 Gyr ago cannot explain the observed very old  low-$\alpha$ \citep{Borbolato2025}, which are instead explained by our  revised parallel model. Thus, it is of fundamental importance to take into account also the age uncertainties and in the following we will consider also the observational errors in our analysis.
\subsection{Model results taking into account the
observational errors}
\begin{figure}
\centering
 	\includegraphics[scale=0.25]{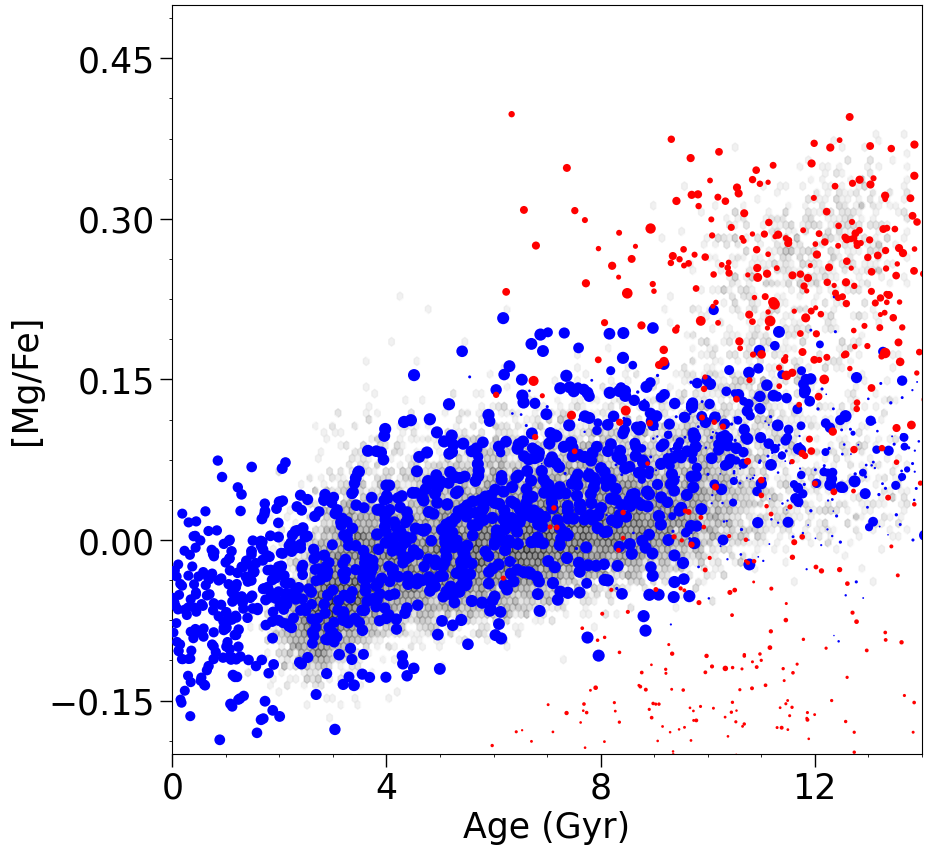}
    \caption{Observed and predicted [$\alpha$/Fe] vs. age (Gyr). Data are from the final sample of \citet[in grey]{Borbolato2025} and the predictions are for the revised parallel model taking into account observational errors (blue dots correspond to the thin disc model, red dots to the thick discl). The size of the coloured points is proportional to the number of stars formed in the corresponding simple stellar population at that Galactic age.}
    \label{MgFe_age_errors}
\end{figure}

In Fig. \ref{MgFe_age_errors}, we show the [$\alpha$/Fe] vs. age  diagram for the sample of B25 compared to predictions of the revised model taking into account observational errors (blue dots correspond to the chemical evolution model of the thin disc, red dots to the thick disc). 
By following \cite{Spitoni2019,Spitoni2023} we consider, a posteriori, the dispersion in the abundance ratios and ages for the
predicted simple stellar populations (SSPs). At each Galactic time, we add a random error to the ages and abundance ratios ([X/H]) of the SSPs formed at Galactic evolutionary time t as follows:
\begin{equation}  \label{eq_age}
\text{Age}_{\text{new}}(t) = \text{Age}(t) + \delta_G(\text{Age}), \quad \delta_G(\text{Age}) \sim \mathcal{N}(0, \sigma_{\text{Age}})
\end{equation}
where $\delta_G$ is a perturbation that follows a normal distribution N(0, $\sigma_{\text{Age}}$) with the standard deviation fixed at the value of $\sigma_{\text{Age}}$ = 20$\%$ Age, compatible with \cite{Queiroz2023}. By analysing \cite{Queiroz2023} data, B25 clearly identified low-$\alpha$ stars with dynamically cooler orbits and ages exceeding 11 Gyr. Such old low-$\alpha$ stars were not so evident in the APOKASC data by \cite{SilvaAguirre2018} used in \cite{Spitoni2019}; with their large uncertainities for the oldest stars, it was not possible to clearly identify the presence of old low-$\alpha$ stars as an effective population and not an artifact of the large uncertainities in their age determination, as instead done in B25.
\\Similarly to Eq. \ref{eq_age} for the stellar ages, for the chemical abundances, we have that:
\begin{equation}
\begin{aligned}
[\text{Mg/Fe}]_{\text{new}}(t) &= [\text{Mg/Fe}](t) + \delta_G([\text{Mg/Fe}]), \\
\delta_G([\text{Mg/Fe}]) &\sim \mathcal{N}(0, \sigma_{[\text{Mg/Fe}]})
\end{aligned}
\end{equation}
where we imposed that $\sigma_{[\text{Mg/Fe}]}$ = 0.05 dex.
From Fig. \ref{MgFe_age_errors}, we can see that we can explain the old stars in the low-$\alpha$ sequence as old thin disc stars (blue dots). Alternatively, old low-$\alpha$ stars can also be interpreted as the metal-rich tail of the thick disc (red dots), even if there are very few low-$\alpha$ thick disc stars.
Finally, we note that different age samples could require different assumptions for the time of formation of the various disc components, since the effective timing of formation of the Galactic disc is still a hot topic in Galactic archaeology (see \citealt{Zhang2024} and references therein). Further data that can precisely fix the beginning of the disc formation would be fundamental to constrain the parameters of our chemical evolution models and high-precision data would be required to put strong constraints on the early phases of the Galaxy evolution (see e.g. \citealt{Montalban2021}). In this context, future missions with high-precision asteroseismology such as HAYDN \citep{haydn2021} could be fundamental.
\\To summarize, a revised parallel model with a pre-enriched and delayed second infall episode, allows us to explain the chemical evolution in the Milky Way high- and low-$\alpha$ sequences and their age distributions. In particular, it can account for i) a short phase of co-evolution between the two components and ii) the existence of very old stars in the low-$\alpha$ sequence. However, we note that the exact onset of the Galactic disc formation still remains to be precisely determined and further data might be necessary to impose more stringent constraints on the parameters of our chemical evolution models. In this context, the revised parallel scenario offers a unique opportunity to follow in detail the evolution of the Galactic thick and thin discs in time and chemical abundance space.


\begin{figure*}
\centering
 	\includegraphics[scale=0.5]{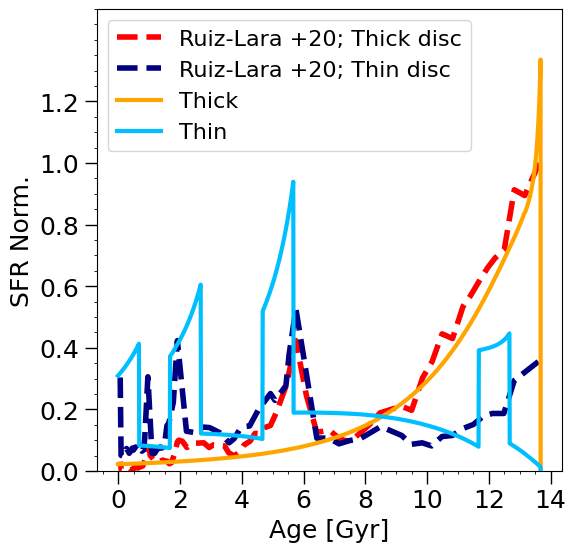}\\
        \includegraphics[scale=0.3]{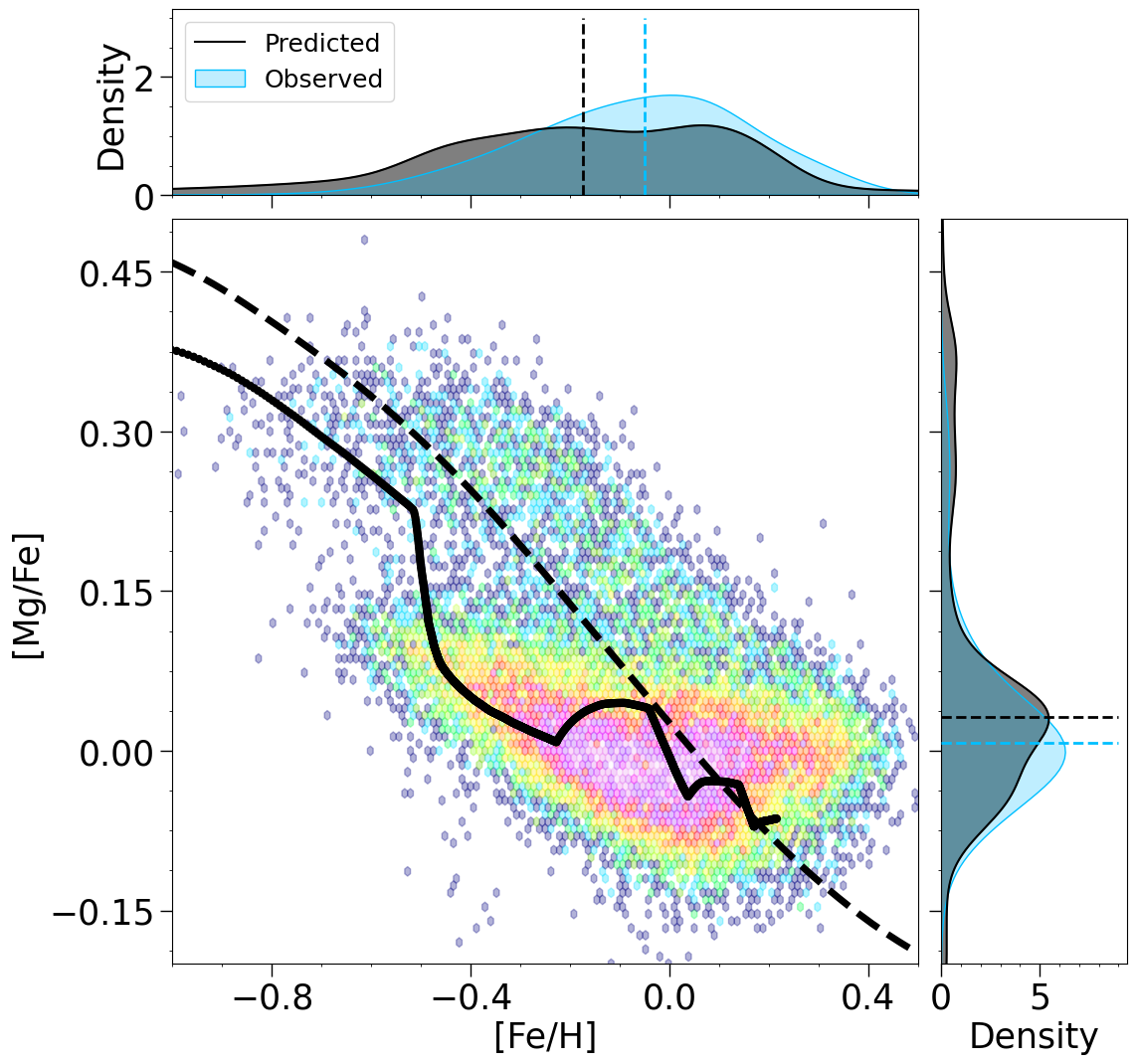}  
                \includegraphics[scale=0.3]{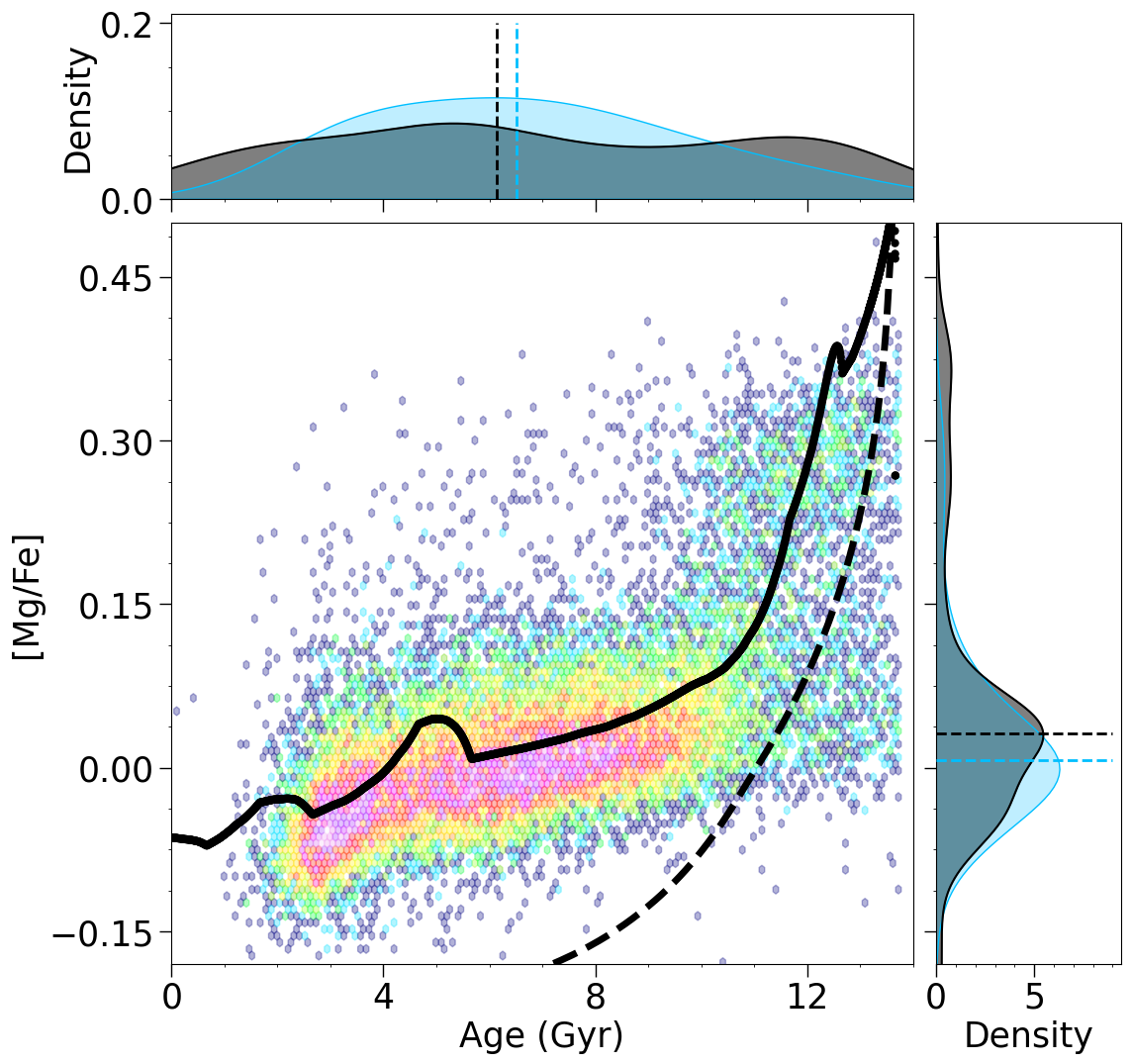}  
    \caption{Upper panel: SFH for our parallel model with bursts compared to the SFHs of \cite{Ruiz-Lara2020} for the thick and thin discs. Lower panel: [$\alpha$/Fe] vs. [Fe/H] (left) and vs. age (right). Data are from the final sample of \cite{Borbolato2025} with chemical abundances from APOGEE DR17 and stellar ages from \flag{StarHorse}. On the sides of each panel, the observed (light blue shaded area) and predicted (dark grey shaded area) normalised KDEs of the distributions calculated with a Gaussian kernel are reported, together with their corresponding medians.
    }
    \label{RuizLara}
\end{figure*}

\subsection{Results for the star formation history of \citet{Ruiz-Lara2020}}

As mentioned in the Introduction, \cite{Ruiz-Lara2020} suggested the SFHs of kinematically selected thick and thin discs, showing a parallel evolution as in the parallel chemical evolution model of \cite{Grisoni2017}. The SFHs by \cite{Ruiz-Lara2020} are shown in Fig. \ref{RuizLara}, together with the predictions of the parallel chemical evolution model, where we imposed also similar bursts of star formation in the thin disc, as observationally indicated. The different starbursts at $\sim$ 6, 2 and 1 Gyr ago in the Galactic disc are interpreted by those authors as due to the passages of Sagittarius galaxy. Given these star formation histories, we implement them in our parallel chemical evolution scheme.
\\In the lower panels of Fig. \ref{RuizLara}, we show the comparison for our parallel chemical evolution model with SFHs following \cite{Ruiz-Lara2020} to observational data, both for abundance patterns and stellar ages. In the lower left panel of Fig. \ref{RuizLara}, we show the predicted [Mg/Fe] vs. [Fe/H] relation for our parallel model with SFHs following \cite{Ruiz-Lara2020} compared to our APOGEE DR17 sample. As we can see from this figure, the behaviour is very similar to the classical parallel model of \cite{Grisoni2017}, with both discs starting forming at the same time a long time ago and proceeding in parallel. The thick disc is $\alpha$-enhanced, due to its higher star formation efficiency and shorter timescale of formation with respect to the thin disc, which on the other hand forms most of these stars later on and suffers subsequent bursts, as suggested observationally. We found that  these star formation histories can reproduce the abundance patterns also with the inclusion of star-formation bursts, even if they would likely be indistinguishable in the chemical properties of stars. 
The two predicted chemical sequences (high-$\alpha$ and low-$\alpha$ ones) eventually merge at [Fe/H]$\sim$0.2 dex (see also \citealt{Hayden2015}). The MDF is reasonably reproduced. A bimodality in the [Mg/Fe] distributions can be also seen, as due to the lag between the first bursts of star formation and the subsequent ones in the thin disc (at variance with the classical parallel model where  star formation is almost continuous). Having subsequent bursts can produce a more peaked [Mg/Fe] distributions in the thin disc and improve then the [Mg/Fe] distributions with respect to the classical parallel model.
\\In the lower right panel of Fig. \ref{RuizLara}, we show the predicted [$\alpha$/Fe] vs. age relation for our parallel model with SFHs following \cite{Ruiz-Lara2020} compared to the sample of B25. Here, we encounter a situation similar to the classical parallel model, with many old stars predicted in the age distributions, even if better than the classical parallel model since we do not have the bulk of the thin disc at very old ages but the thin disc forms most of its stars later on during the  subsequent bursts of star formation, and at the beginning we have only a proto-thin disc. In this case, however, without the pre-enrichment for the low-$\alpha$ sequence, the low-$\alpha$ curve still starts from high [Mg/Fe] values, at variance with observations. This is similar to the classical parallel case, and we have shown in this paper, that this situation can be solved by adding some pre-enrichment level and a short delay for the second infall episode, with the low-$\alpha$ curve then actually starting in the low-$\alpha$ region.
For the thick disc curve, we have a similar situation as in the classical parallel case. However, to best match the age distributions of thick disc stars we have shown that a more truly old population is needed,  obtained with a much faster evolution than in this case.
In summary, the proposed history of star formation by \cite{Ruiz-Lara2020} produces a  reasonable fit of the [$\alpha$/Fe] vs. [Fe/H] diagram with two sequences corresponding to the thick-disc ($\alpha$-enhanced) and thin-disc (low-$\alpha$) stars. However, it encounters similar issues to explain the [Mg/Fe] vs age plots as the classical parallel model, with the two discs starting forming exactly at the same time $\sim$ 13 Gyr ago with high [Mg/Fe] value. We then conclude that our proposed new parallel model better describes the evolution of the two discs.
\\Finally, we note that recent bursts in the SFH as proposed by \cite{Ruiz-Lara2020} can enhance the [Mg/Fe] values at young ages; this has been tested in \cite{Prantzos2025}  but we have done it here by means of a parallel approach that can follow separately the evolution in the two discs. 
We also find that the presence of bursts can alleviate  but still not solve the problem of young-$\alpha$ rich stars in the Galaxy \citep{Chiappini2015,Martig2015,Cerqui2023,Grisoni2024}. To reach the young $\alpha$-rich stars, much stronger bursts would be needed (ten times higher than assumed here). However, we know that young $\alpha$-rich stars might not be truly young, but rather products of binary evolution and mass transfer (\citealt{Grisoni2024} and references therein). To conclude, the analysis suggests that even if star-formation bursts occurred, they would likely be indistinguishable in the chemical properties of stars, given current data quality. As such, the modelling neither confirms nor robustly rules out their presence.

\subsection{Discussion}

To summarize, in a parallel chemical evolution approach, the two discs form in parallel but at different rates. This approach is supported also by the work of \cite{Clarke2019} where the two distinct sequences can be explained as disc fragmentation: the high-$\alpha$ sequence is related to clumps with high star formation efficiency, while more distributed and less efficient star formation produces the low-$\alpha$ sequence. Another interpretation of the bimodality relies on the idea of a late second gas accretion episode after a period of quenched star formation like has been proposed in the works of \cite{Grand2018} and \cite{Mackereth2018}. \cite{Buck2020} explain the bimodal $\alpha$-sequence as a generic consequence of a gas-rich merger at some time in the Galaxy's evolution: the high-$\alpha$ sequence evolves first in the early galaxy extending to high metallicities, while it is the low-$\alpha$ sequence that is formed after the gas-rich merger (see also \citealt{Calura2009}). Another interpretation for the formation of the bimodality in the [$\alpha$/Fe] vs. [Fe/H] plot is associated to radial migration (\citealt{Schoenrich2009} and more recently \citealt{Sharma2021,Prantzos2023}), even if \cite{Minchev2013} found a smooth density distribution in the [O/Fe]-[Fe/H] plane.  \cite{Palla2022} tested radial migration in both one-infall and two-infall chemical evolution models, but they showed that they have a small effect on the
overall distribution functions, and that peculiar histories
of star formation are still needed to explain the distribution of stars in the Galaxy \citep{Mackereth2018,Vincenzo2021,Khoperskov2021}.
In general, to reproduce a bimodality in [$\alpha$/Fe] vs. [Fe/H], we propose a very short timescale of formation for the thick disc and a much longer evolution for the thin disc, and a period of low star formation between the peaks of the two phases (see also the quenching in the star formation history of the Milky Way found by \citealt{Haywood2016}). This reconciles with the idea of a hiatus in star formation as proposed in \cite{Spitoni2024}. Moreover, we can explain also a phase of co-evolution between the two disc components, as recently proposed \citep{Beraldo2021,Gent2024}. Also, in our scenario we can interpret very old stars in the low-$\alpha$ sequence, as recently found by B25. To conclude, our chemical evolution models offer a unique opportunity to follow in detail the evolution in the Milky Way thick and thin discs, and can also naturally explain: i) a co-evolution phase between the two disc components, ii) old low-$\alpha$ stars, as found by recent observational data.


\section{Summary and conclusions}

We have studied the chemical evolution of the Milky Way thick and thin discs by comparing detailed chemical evolution models to recent observational data. We have started from the original parallel chemical evolution approach of \cite{Grisoni2017}, i.e. two distinct one-infall models for the Galactic thick and thin discs. 
Very recent observational studies \citep[see e.g.][and references therein]{Borbolato2025} have indicated the existence of very old low-$\alpha$ stars, invoking the need for a parallel chemical evolution approach. Moreover, the parallel approach can account for a phase of co-evolution among the two disc components \citep{Beraldo2021,Gent2024}. However, \cite{Spitoni2019} showed that two discs starting forming exactly at the same time could be problematic to explain the stellar age distribution of APOKASC. Therefore, we revised the parallel approach, by considering pre-enriched and delayed infall of gas in order to best reproduce the most recent observational data for the high- and low-$\alpha$ disc stars. In addition, we have tested the effects of the star formation histories derived by \citet{Ruiz-Lara2020} for kinematically selected thick and thin discs, showing a parallel evolution for the two components and the presence of bursts of star formation.\\

Our conclusions, based on the comparison between our chemical evolution models and the selected sample from \cite{Borbolato2025}, can be summarized as follows.
\begin{itemize}
\item Both the classical parallel model and the revised model with  a pre-enriched delayed second infall episode can explain the two sequences  in the [$\alpha$/Fe] vs. [Fe/H] relation  observed in the thick and thin discs (high- and low-$\alpha$ sequences, respectively) from APOGEE DR17 data. However, to have a clear bimodality in the [Mg/Fe] distributions, we need a very old thick disc coupled with  a pre-enriched delayed second infall episode for creating the low-$\alpha$ sequence, and this is obtained with the revised model. By considering pre-enriched and delayed infall of gas, the revised scenario gets closer to a two-infall case, still with the ability to follow the two chemical sequences separately.
\item Our revised model can also explain the age distributions for the selected observational sample from \cite{Borbolato2025}, at variance with the classical parallel approach where primordial infall and no delay were considered and there was a bulk of old low-$\alpha$ stars. 
\item As also motivated by various stellar age studies \citep{Miglio2021,Queiroz2023}, in our scenario the thick disc forms very fast (less than a billion years), it is a very old population of stars with ages peaked at $\sim$12 Gyr ago. On the other hand, the thin disc forms much more slowly (several Gyrs in the solar vicinity).
\item A hiatus in the star formation (see also \citealt{Haywood2016}) can emerge as a period of low star formation between the formation of the high-$\alpha$ sequence (forming fast, short timescale) and the low-$\alpha$ one (forming on a longer timescale of formation).
\item We predict a co-evolution phase for the thick and thin discs, as recently observed by several studies \citep{Beraldo2021,Wu2023,Gent2024}. However, at variance with the classical parallel of \cite{Grisoni2017}, the revised model features a shorter overlap between the formation of the high- and low-$\alpha$ sequence. 
\item In the framework of the revised model including also the observational uncertainties, we can explain the presence of old low-$\alpha$ stars, as recently observed  by \cite{Borbolato2025}. In this context, the model provides a plausible explanation for those old low-$\alpha$ stars if the adopted age scale is correct, but still further observational data would be needed to put strong constraints on the early phases of the Galactic evolution and the physical nature of those old low-$\alpha$ stars.
\item Finally, we tested also observationally derived star formation histories of the kinematically selected thick and thin disc as suggested by \cite{Ruiz-Lara2020}, that show a parallel evolution for these two components and the presence of star-formation bursts. We found that  these star formation histories can reproduce the abundance patterns also with the inclusion of star-formation bursts, even if they would likely be indistinguishable in the chemical properties of stars.  The star formation rate for the thin disc can well reproduce the data also with the inclusion of bursts; on the other hand, a prolonged star formation history for the thick disc is not compatible with its observed stellar age distribution of a very old population of stars.
\end{itemize}
In summary, Galactic chemical evolution models are fundamental tools to follow in detail the chemical evolution of our home Milky Way in comparison with the data for chemical abundances and stellar ages. In particular, our proposed chemical evolution scenario can explain i) a co-evolution phase between the Galactic high- and low-$\alpha$ sequences, and ii) the presence of old stars in the low-$\alpha$ sequence, as in the selected observational sample. Nevertheless, further observational data are really fundamental to put more stringent constraints on the timescale for the formation of the Galactic disc, in particular in the earliest Galactic phases, which still represent a hot topic in Galactic archaeology (see e.g. \citealt{Zhang2024} and references therein) To this purpose, the high-precision of HAYDN \citep{haydn2021} could be fundamental. Other facilities such as a Wide-field Spectroscopic Telescope (WST, \citealt{WST2024}) can help to constrain the formation timescale of the Galactic disc (in particular see \citealt{WST2025}).

\begin{acknowledgements}
We are grateful to the referee for the careful reading and all the very useful comments and suggestions, which have improved the presentation of our work. We thank Lais Borbolato for kindly sharing the final data sample from the paper \cite{Borbolato2025}. V.G. acknowledges financial support from INAF under the program “Giovani Astrofisiche ed Astrofisici di Eccellenza - IAF: INAF Astrophysics Fellowships in Italy" (Project: GalacticA, "Galactic Archaeology: reconstructing the history of the Galaxy") and Minigrant INAF 2023. V.G. acknowledges useful conversations with Cristina Chiappini. E.S. and F.M.   thank I.N.A.F. for the  
1.05.24.07.02 Mini Grant - LEGARE "Linking the chemical Evolution of Galactic discs AcRoss diversE scales: from the thin disc to the nuclear stellar disc" (PI E. Spitoni).
\end{acknowledgements}


\bibliographystyle{aa}
\bibliography{aa_biblio}

@ARTICLE{Argast2000,
       author = {{Argast}, D. and {Samland}, M. and {Gerhard}, O.~E. and {Thielemann}, F.-K.},
        title = "{Metal-poor halo stars as tracers of ISM mixing processes during halo formation}",
      journal = {\aap},
         year = 2000,
        month = apr,
       volume = {356},
        pages = {873-887},
          doi = {10.48550/arXiv.astro-ph/9911178},
archivePrefix = {arXiv},
       eprint = {astro-ph/9911178},
 primaryClass = {astro-ph},
       adsurl = {https://ui.adsabs.harvard.edu/abs/2000A&A...356..873A}
}

@ARTICLE{Snaith2015,
       author = {{Snaith}, O. and {Haywood}, M. and {Di Matteo}, P. and {Lehnert}, M.~D. and {Combes}, F. and {Katz}, D. and {G{\'o}mez}, A.},
        title = "{Reconstructing the star formation history of the Milky Way disc(s) from chemical abundances}",
      journal = {\aap},
         year = 2015,
        month = jun,
       volume = {578},
          eid = {A87},
        pages = {A87},
          doi = {10.1051/0004-6361/201424281},
archivePrefix = {arXiv},
       eprint = {1410.3829},
 primaryClass = {astro-ph.GA},
       adsurl = {https://ui.adsabs.harvard.edu/abs/2015A&A...578A..87S}
}

@ARTICLE{WST2025,
       author = {{Bergemann}, M. and {Kordopatis}, G. and {Casali}, G. and {Khoperskov}, S. and {McMillan}, P. and {Marques}, L. and {Minchev}, I. and {Poggio}, E. and {Schultheis}, M. and {Viscasillas V{\'a}zquez}, C. and {Wang}, H.-F. and {Grisoni}, V. and {Hill}, V. and {Smiljanic}, R.},
        title = "{Milky Way disc \& Bulge in situ populations: ESO white paper - Expanding horizons call}",
      journal = {arXiv e-prints},
         year = 2025,
        month = dec,
          eid = {arXiv:2512.15812},
        pages = {arXiv:2512.15812},
          doi = {10.48550/arXiv.2512.15812},
archivePrefix = {arXiv},
       eprint = {2512.15812},
 primaryClass = {astro-ph.IM},
       adsurl = {https://ui.adsabs.harvard.edu/abs/2025arXiv251215812B}
}

@ARTICLE{Alinder2025,
       author = {{Alinder}, Simon and {Bensby}, Thomas and {McMillan}, Paul},
        title = "{Impact of selection criteria on the structural parameters of the Galactic thin and thick discs}",
      journal = {arXiv e-prints},
         year = 2025,
        month = nov,
          eid = {arXiv:2511.10092},
        pages = {arXiv:2511.10092},
          doi = {10.48550/arXiv.2511.10092},
archivePrefix = {arXiv},
       eprint = {2511.10092},
 primaryClass = {astro-ph.GA},
       adsurl = {https://ui.adsabs.harvard.edu/abs/2025arXiv251110092A}
}

@ARTICLE{WST2024,
       author = {{Mainieri}, Vincenzo and {Anderson}, Richard I. and {Brinchmann}, Jarle and {Cimatti}, Andrea and {Ellis}, Richard S. and {Hill}, Vanessa and {Kneib}, Jean-Paul and {McLeod}, Anna F. and {Opitom}, Cyrielle and {Roth}, Martin M. and {Sanchez-Saez}, Paula and {Smiljanic}, Rodolfo and {Tolstoy}, Eline and {Bacon}, Roland and {Randich}, Sofia and {Adamo}, Angela and {Annibali}, Francesca and {Arevalo}, Patricia and {Audard}, Marc and {Barsanti}, Stefania and {Battaglia}, Giuseppina and {Bayo Aran}, Amelia M. and {Belfiore}, Francesco and {Bellazzini}, Michele and {Bellini}, Emilio and {Beltran}, Maria Teresa and {Berni}, Leda and {Bianchi}, Simone and {Biazzo}, Katia and {Bisero}, Sofia and {Bisogni}, Susanna and {Bland-Hawthorn}, Joss and {Blondin}, Stephane and {Bodensteiner}, Julia and {Boffin}, Henri M.~J. and {Bonito}, Rosaria and {Bono}, Giuseppe and {Bouche}, Nicolas F. and {Bowman}, Dominic and {Braga}, Vittorio F. and {Bragaglia}, Angela and {Branchesi}, Marica and {Brucalassi}, Anna and {Bryant}, Julia J. and {Bryson}, Ian and {Busa}, Innocenza and {Camera}, Stefano and {Carbone}, Carmelita and {Casali}, Giada and {Casali}, Mark and {Casasola}, Viviana and {Castro}, Norberto and {Catelan}, Marcio and {Cavallo}, Lorenzo and {Chiappini}, Cristina and {Cioni}, Maria-Rosa and {Colless}, Matthew and {Colzi}, Laura and {Contarini}, Sofia and {Couch}, Warrick and {D'Ammando}, Filippo and {d'Assignies D.}, William and {D'Orazi}, Valentina and {da Silva}, Ronaldo and {Dainotti}, Maria Giovanna and {Damiani}, Francesco and {Danielski}, Camilla and {De Cia}, Annalisa and {de Jong}, Roelof S. and {Dhawan}, Suhail and {Dierickx}, Philippe and {Driver}, Simon P. and {Dupletsa}, Ulyana and {Escoffier}, Stephanie and {Escorza}, Ana and {Fabrizio}, Michele and {Fiorentino}, Giuliana and {Fontana}, Adriano and {Fontani}, Francesco and {Forero Sanchez}, Daniel and {Franois}, Patrick and {Galindo-Guil}, Francisco Jose and {Gallazzi}, Anna Rita and {Galli}, Daniele and {Garcia}, Miriam and {Garcia-Rojas}, Jorge and {Garilli}, Bianca and {Grand}, Robert and {Guarcello}, Mario Giuseppe and {Hazra}, Nandini and {Helmi}, Amina and {Herrero}, Artemio and {Iglesias}, Daniela and {Ilic}, Dragana and {Irsic}, Vid and {Ivanov}, Valentin D. and {Izzo}, Luca and {Jablonka}, Pascale and {Joachimi}, Benjamin and {Kakkad}, Darshan and {Kamann}, Sebastian and {Koposov}, Sergey and {Kordopatis}, Georges and {Kovacevic}, Andjelka B. and {Kraljic}, Katarina and {Kuncarayakti}, Hanindyo and {Kwon}, Yuna and {La Forgia}, Fiorangela and {Lahav}, Ofer and {Laigle}, Clotilde and {Lazzarin}, Monica and {Leaman}, Ryan and {Leclercq}, Floriane and {Lee}, Khee-Gan and {Lee}, David and {Lehnert}, Matt D. and {Lira}, Paulina and {Loffredo}, Eleonora and {Lucatello}, Sara and {Magrini}, Laura and {Maguire}, Kate and {Mahler}, Guillaume and {Zahra Majidi}, Fatemeh and {Malavasi}, Nicola and {Mannucci}, Filippo and {Marconi}, Marcella and {Martin}, Nicolas and {Marulli}, Federico and {Massari}, Davide and {Matsuno}, Tadafumi and {Mattheee}, Jorryt and {McGee}, Sean and {Merc}, Jaroslav and {Merle}, Thibault and {Miglio}, Andrea and {Migliorini}, Alessandra and {Minchev}, Ivan and {Minniti}, Dante and {Miret-Roig}, Nuria and {Monreal Ibero}, Ana and {Montano}, Federico and {Montet}, Ben T. and {Moresco}, Michele and {Moretti}, Chiara and {Moscardini}, Lauro and {Moya}, Andres and {Mueller}, Oliver and {Nanayakkara}, Themiya and {Nicholl}, Matt and {Nordlander}, Thomas and {Onori}, Francesca and {Padovani}, Marco and {Pala}, Anna Francesca and {Panda}, Swayamtrupta and {Pandey-Pommier}, Mamta and {Pasquini}, Luca and {Pawlak}, Michal and {Pessi}, Priscila J. and {Pisani}, Alice and {Popovic}, Lukav C. and {Prisinzano}, Loredana and {Raddi}, Roberto and {Rainer}, Monica and {Rebassa-Mansergas}, Alberto and {Richard}, Johan and {Rigault}, Mickael and {Rocher}, Antoine and {Romano}, Donatella and {Rosati}, Piero and {Sacco}, Germano and {Sanchez-Janssen}, Ruben and {Sander}, Andreas A.~C. and {Sanders}, Jason L. and {Sargent}, Mark and {Sarpa}, Elena and {Schimd}, Carlo and {Schipani}, Pietro and {Sefusatti}, Emiliano and {Smith}, Graham P. and {Spina}, Lorenzo and {Steinmetz}, Matthias and {Tacchella}, Sandro and {Tautvaisiene}, Grazina and {Theissen}, Christopher and {Thomas}, Guillaume and {Ting}, Yuan-Sen and {Travouillon}, Tony and {Tresse}, Laurence and {Trivedi}, Oem and {Tsantaki}, Maria and {Tsedrik}, Maria and {Urrutia}, Tanya and {Valenti}, Elena and {Van der Swaelmen}, Mathieu and {Van Eck}, Sophie and {Verdiani}, Francesco and {Verdier}, Aurelien and {Vergani}, Susanna Diana and {Verhamme}, Anne and {Vernet}, Joel},
        title = "{The Wide-field Spectroscopic Telescope (WST) Science White Paper}",
      journal = {arXiv e-prints},
         year = 2024,
        month = mar,
          eid = {arXiv:2403.05398},
        pages = {arXiv:2403.05398},
          doi = {10.48550/arXiv.2403.05398},
archivePrefix = {arXiv},
       eprint = {2403.05398},
 primaryClass = {astro-ph.IM},
       adsurl = {https://ui.adsabs.harvard.edu/abs/2024arXiv240305398M}
}

@ARTICLE{Casali2025,
       author = {{Casali}, G. and {Montalb{\'a}n}, J. and {Miglio}, A. and {Casagrande}, L. and {Magrini}, L. and {Chiappini}, C. and {Bragaglia}, A. and {Matteuzzi}, M. and {Brogaard}, K. and {Stokholm}, A. and {Grisoni}, V. and {Tailo}, M. and {Willett}, E.},
        title = "{Tracing the Milky Way: calibrating chemical ages with high-precision Kepler data}",
      journal = {\mnras},
         year = 2025,
        month = aug,
       volume = {541},
       number = {3},
        pages = {2631-2650},
          doi = {10.1093/mnras/staf1047},
archivePrefix = {arXiv},
       eprint = {2506.15546},
 primaryClass = {astro-ph.GA},
       adsurl = {https://ui.adsabs.harvard.edu/abs/2025MNRAS.541.2631C}
}

@ARTICLE{haydn2021,
       author = {{Miglio}, Andrea and {Girardi}, L{\'e}o and {Grundahl}, Frank and {Mosser}, Benoit and {Bastian}, Nate and {Bragaglia}, Angela and {Brogaard}, Karsten and {Buldgen}, Ga{\"e}l and {Chantereau}, William and {Chaplin}, William and {Chiappini}, Cristina and {Dupret}, Marc-Antoine and {Eggenberger}, Patrick and {Gieles}, Mark and {Izzard}, Robert and {Kawata}, Daisuke and {Karoff}, Christoffer and {Lagarde}, Nad{\`e}ge and {Mackereth}, Ted and {Magrin}, Demetrio and {Meynet}, Georges and {Michel}, Eric and {Montalb{\'a}n}, Josefina and {Nascimbeni}, Valerio and {Noels}, Arlette and {Piotto}, Giampaolo and {Ragazzoni}, Roberto and {Soszy{\'n}ski}, Igor and {Tolstoy}, Eline and {Toonen}, Silvia and {Triaud}, Amaury and {Vincenzo}, Fiorenzo},
        title = "{Haydn}",
      journal = {Experimental Astronomy},
         year = 2021,
        month = jun,
       volume = {51},
       number = {3},
        pages = {963-1001},
          doi = {10.1007/s10686-021-09711-1},
archivePrefix = {arXiv},
       eprint = {1908.05129},
 primaryClass = {astro-ph.SR},
       adsurl = {https://ui.adsabs.harvard.edu/abs/2021ExA....51..963M}
}

@ARTICLE{Kubryk2013,
       author = {{Kubryk}, M. and {Prantzos}, N. and {Athanassoula}, E.},
        title = "{Radial migration in a bar-dominated disc galaxy - I. Impact on chemical evolution}",
      journal = {\mnras},
         year = 2013,
        month = dec,
       volume = {436},
       number = {2},
        pages = {1479-1491},
          doi = {10.1093/mnras/stt1667},
       adsurl = {https://ui.adsabs.harvard.edu/abs/2013MNRAS.436.1479K}
}

@ARTICLE{Gallart2024,
       author = {{Gallart}, Carme and {Surot}, Francisco and {Cassisi}, Santi and {Fern{\'a}ndez-Alvar}, Emma and {Mirabal}, David and {Rivero}, Alicia and {Ruiz-Lara}, Tom{\'a}s and {Santos-Torres}, Judith and {Aznar-Menargues}, Guillem and {Battaglia}, Giuseppina and {Queiroz}, Anna B. and {Monelli}, Matteo and {Vasiliev}, Eugene and {Chiappini}, Cristina and {Helmi}, Amina and {Hill}, Vanessa and {Massari}, Davide and {Thomas}, Guillaume F.},
        title = "{Chronology of our Galaxy from Gaia colour-magnitude diagram fitting (ChronoGal). I. The formation and evolution of the thin disc from the Gaia Catalogue of Nearby Stars}",
      journal = {\aap},
         year = 2024,
        month = jul,
       volume = {687},
          eid = {A168},
        pages = {A168},
          doi = {10.1051/0004-6361/202349078},
archivePrefix = {arXiv},
       eprint = {2402.09399},
 primaryClass = {astro-ph.GA},
       adsurl = {https://ui.adsabs.harvard.edu/abs/2024A&A...687A.168G}
}

@ARTICLE{Dubay2025,
       author = {{Dubay}, Liam O. and {Johnson}, Jennifer A. and {Johnson}, James W. and {Roberts}, John D.},
        title = "{Challenges to the Two-Infall Scenario by Large Stellar Age Catalogs}",
      journal = {arXiv e-prints},
         year = 2025,
        month = aug,
          eid = {arXiv:2508.00988},
        pages = {arXiv:2508.00988},
          doi = {10.48550/arXiv.2508.00988},
archivePrefix = {arXiv},
       eprint = {2508.00988},
 primaryClass = {astro-ph.GA},
       adsurl = {https://ui.adsabs.harvard.edu/abs/2025arXiv250800988D}
}

@ARTICLE{Zhang2025,
       author = {{Zhang}, HanYuan and {Iorio}, Giuliano and {Belokurov}, Vasily and {Evans}, N. Wyn and {Bobrick}, Alexey and {D'Orazi}, Valentina},
        title = "{Revealing the ages of metal-rich RR Lyrae via kinematic label transfer}",
      journal = {arXiv e-prints},
         year = 2025,
        month = apr,
          eid = {arXiv:2504.06720},
        pages = {arXiv:2504.06720},
          doi = {10.48550/arXiv.2504.06720},
archivePrefix = {arXiv},
       eprint = {2504.06720},
 primaryClass = {astro-ph.SR},
       adsurl = {https://ui.adsabs.harvard.edu/abs/2025arXiv250406720Z}
}

@ARTICLE{Borbolato2025,
       author = {{Borbolato}, Lais and {Rossi}, Silvia and {Perottoni}, H{\'e}lio D. and {Limberg}, Guilherme and {Amarante}, Jo{\~a}o A.~S. and {Queiroz}, Anna B.~A. and {Chiappini}, Cristina and {Anders}, Friedrich and {Santucci}, Rafael M. and {Barbosa}, Fabr{\'\i}cia O. and {Nogueira-Santos}, Jo{\~a}o V.},
        title = "{Early Coformation of the Milky Way's Thin and Thick Disks at Redshift z > 2}",
      journal = {\apj},
         year = 2025,
        month = nov,
       volume = {994},
       number = {1},
          eid = {126},
        pages = {126},
          doi = {10.3847/1538-4357/ae0c96},
archivePrefix = {arXiv},
       eprint = {2504.00135},
 primaryClass = {astro-ph.GA},
       adsurl = {https://ui.adsabs.harvard.edu/abs/2025ApJ...994..126B}
}

@ARTICLE{Cescutti2020,
       author = {{Cescutti}, G. and {Molaro}, P. and {Fu}, X.},
        title = "{Lithium in the closest satellite of our Milky Way}",
      journal = {\memsai},
         year = 2020,
        month = jan,
       volume = {91},
        pages = {153},
          doi = {10.48550/arXiv.2004.06606},
archivePrefix = {arXiv},
       eprint = {2004.06606},
 primaryClass = {astro-ph.SR},
       adsurl = {https://ui.adsabs.harvard.edu/abs/2020MmSAI..91..153C}
}

@ARTICLE{Cescutti2008,
       author = {{Cescutti}, G.},
        title = "{An inhomogeneous model for the Galactic halo: a possible explanation for the spread observed in s- and r-process elements}",
      journal = {\aap},
         year = 2008,
        month = apr,
       volume = {481},
       number = {3},
        pages = {691-699},
          doi = {10.1051/0004-6361:20078571},
archivePrefix = {arXiv},
       eprint = {0802.0678},
 primaryClass = {astro-ph},
       adsurl = {https://ui.adsabs.harvard.edu/abs/2008A&A...481..691C}
}

@ARTICLE{Willett2026,
       author = {{Willett}, Emma and {Miglio}, Andrea and {Khan}, Saniya and {Elsworth}, Yvonne and {Mosser}, Beno{\^\i}t and {Brogaard}, Karsten and {Casali}, Giada and {Chiappini}, Cristina and {Grisoni}, Valeria and {Stokholm}, Amalie and {Bossini}, Diego and {Chaplin}, William J.},
        title = "{Asteroseismic ages for 17,000 stars in Kepler, K2 and TESS}",
      journal = {arXiv e-prints},
         year = 2026,
        month = feb,
          eid = {arXiv:2602.06870},
        pages = {arXiv:2602.06870},
archivePrefix = {arXiv},
       eprint = {2602.06870},
 primaryClass = {astro-ph.SR},
       adsurl = {https://ui.adsabs.harvard.edu/abs/2026arXiv260206870W}
}

@ARTICLE{smith2021,
       author = {{Smith}, Verne V. and {Bizyaev}, Dmitry and {Cunha}, Katia and {Shetrone}, Matthew D. and {Souto}, Diogo and {Allende Prieto}, Carlos and {Masseron}, Thomas and {M{\'e}sz{\'a}ros}, Szabolcs and {J{\"o}nsson}, Henrik and {Hasselquist}, Sten and {Osorio}, Yeisson and {Garc{\'\i}a-Hern{\'a}ndez}, D.~A. and {Plez}, Bertrand and {Beaton}, Rachael L. and {Holtzman}, Jon and {Majewski}, Steven R. and {Stringfellow}, Guy S. and {Sobeck}, Jennifer},
        title = "{The APOGEE Data Release 16 Spectral Line List}",
      journal = {\aj},
         year = 2021,
        month = jun,
       volume = {161},
       number = {6},
          eid = {254},
        pages = {254},
          doi = {10.3847/1538-3881/abefdc},
archivePrefix = {arXiv},
       eprint = {2103.10112},
 primaryClass = {astro-ph.SR},
       adsurl = {https://ui.adsabs.harvard.edu/abs/2021AJ....161..254S}
}

@ARTICLE{jonsson2020,
       author = {{J{\"o}nsson}, Henrik and {Holtzman}, Jon A. and {Allende Prieto}, Carlos and {Cunha}, Katia and {Garc{\'\i}a-Hern{\'a}ndez}, D.~A. and {Hasselquist}, Sten and {Masseron}, Thomas and {Osorio}, Yeisson and {Shetrone}, Matthew and {Smith}, Verne and {Stringfellow}, Guy S. and {Bizyaev}, Dmitry and {Edvardsson}, Bengt and {Majewski}, Steven R. and {M{\'e}sz{\'a}ros}, Szabolcs and {Souto}, Diogo and {Zamora}, Olga and {Beaton}, Rachael L. and {Bovy}, Jo and {Donor}, John and {Pinsonneault}, Marc H. and {Poovelil}, Vijith Jacob and {Sobeck}, Jennifer},
        title = "{APOGEE Data and Spectral Analysis from SDSS Data Release 16: Seven Years of Observations Including First Results from APOGEE-South}",
      journal = {\aj},
         year = 2020,
        month = sep,
       volume = {160},
       number = {3},
          eid = {120},
        pages = {120},
          doi = {10.3847/1538-3881/aba592},
archivePrefix = {arXiv},
       eprint = {2007.05537},
 primaryClass = {astro-ph.GA},
       adsurl = {https://ui.adsabs.harvard.edu/abs/2020AJ....160..120J}
}

@ARTICLE{gustafsson2008,
       author = {{Gustafsson}, B. and {Edvardsson}, B. and {Eriksson}, K. and {J{\o}rgensen}, U.~G. and {Nordlund}, {\r{A}}. and {Plez}, B.},
        title = "{A grid of MARCS model atmospheres for late-type stars. I. Methods and general properties}",
      journal = {\aap},
         year = 2008,
        month = aug,
       volume = {486},
       number = {3},
        pages = {951-970},
          doi = {10.1051/0004-6361:200809724},
archivePrefix = {arXiv},
       eprint = {0805.0554},
 primaryClass = {astro-ph},
       adsurl = {https://ui.adsabs.harvard.edu/abs/2008A&A...486..951G}
}

@ARTICLE{garciaperez2016,
       author = {{Garc{\'\i}a P{\'e}rez}, Ana E. and {Allende Prieto}, Carlos and {Holtzman}, Jon A. and {Shetrone}, Matthew and {M{\'e}sz{\'a}ros}, Szabolcs and {Bizyaev}, Dmitry and {Carrera}, Ricardo and {Cunha}, Katia and {Garc{\'\i}a-Hern{\'a}ndez}, D.~A. and {Johnson}, Jennifer A. and {Majewski}, Steven R. and {Nidever}, David L. and {Schiavon}, Ricardo P. and {Shane}, Neville and {Smith}, Verne V. and {Sobeck}, Jennifer and {Troup}, Nicholas and {Zamora}, Olga and {Weinberg}, David H. and {Bovy}, Jo and {Eisenstein}, Daniel J. and {Feuillet}, Diane and {Frinchaboy}, Peter M. and {Hayden}, Michael R. and {Hearty}, Fred R. and {Nguyen}, Duy C. and {O'Connell}, Robert W. and {Pinsonneault}, Marc H. and {Wilson}, John C. and {Zasowski}, Gail},
        title = "{ASPCAP: The APOGEE Stellar Parameter and Chemical Abundances Pipeline}",
      journal = {\aj},
         year = 2016,
        month = jun,
       volume = {151},
       number = {6},
          eid = {144},
        pages = {144},
          doi = {10.3847/0004-6256/151/6/144},
archivePrefix = {arXiv},
       eprint = {1510.07635},
 primaryClass = {astro-ph.SR},
       adsurl = {https://ui.adsabs.harvard.edu/abs/2016AJ....151..144G}
}

@ARTICLE{Nesti2013,
       author = {{Nesti}, Fabrizio and {Salucci}, Paolo},
        title = "{The Dark Matter halo of the Milky Way, AD 2013}",
      journal = {\jcap},
         year = 2013,
        month = jul,
       volume = {2013},
       number = {7},
          eid = {016},
        pages = {016},
          doi = {10.1088/1475-7516/2013/07/016},
archivePrefix = {arXiv},
       eprint = {1304.5127},
 primaryClass = {astro-ph.GA},
       adsurl = {https://ui.adsabs.harvard.edu/abs/2013JCAP...07..016N}
}

@ARTICLE{Orkney2025,
       author = {{Orkney}, Matthew D.~A. and {Laporte}, Chervin F.~P. and {Grand}, Robert J.~J. and {Springel}, Volker},
        title = "{The Milky Way in context: The formation of galactic discs and chemical sequences from a cosmological perspective}",
      journal = {arXiv e-prints},
         year = 2025,
        month = jun,
          eid = {arXiv:2506.07038},
        pages = {arXiv:2506.07038},
          doi = {10.48550/arXiv.2506.07038},
archivePrefix = {arXiv},
       eprint = {2506.07038},
 primaryClass = {astro-ph.GA},
       adsurl = {https://ui.adsabs.harvard.edu/abs/2025arXiv250607038O}
}

@ARTICLE{Laporte2020a,
       author = {{Laporte}, Chervin F.~P. and {Belokurov}, Vasily and {Koposov}, Sergey E. and {Smith}, Martin C. and {Hill}, Vanessa},
        title = "{Chemo-dynamical properties of the Anticenter Stream: a surviving disc fossil from a past satellite interaction.}",
      journal = {\mnras},
         year = 2020,
        month = feb,
       volume = {492},
       number = {1},
        pages = {L61-L65},
          doi = {10.1093/mnrasl/slz167},
archivePrefix = {arXiv},
       eprint = {1907.10678},
 primaryClass = {astro-ph.GA},
       adsurl = {https://ui.adsabs.harvard.edu/abs/2020MNRAS.492L..61L}
}

@ARTICLE{Orazi2025,
       author = {{D'Orazi}, V. and {Braga}, V. and {Bono}, G. and {Fabrizio}, M. and {Fiorentino}, G. and {Storm}, N. and {Pietrinferni}, A. and {Sneden}, C. and {S{\'a}nchez-Benavente}, M. and {Monelli}, M. and {Sestito}, F. and {J{\"o}nsson}, H. and {Buder}, S. and {Bobrick}, A. and {Iorio}, G. and {Matsunaga}, N. and {Marconi}, M. and {Marengo}, M. and {Mart{\'\i}nez-V{\'a}zquez}, C.~E. and {Mullen}, J. and {Takayama}, M. and {Testa}, V. and {Cusano}, F. and {Crestani}, J.},
        title = "{The elderly among the oldest: new evidence for extremely metal-poor RR Lyrae stars}",
      journal = {\aap},
         year = 2025,
        month = feb,
       volume = {694},
          eid = {A158},
        pages = {A158},
          doi = {10.1051/0004-6361/202453202},
archivePrefix = {arXiv},
       eprint = {2501.05807},
 primaryClass = {astro-ph.SR},
       adsurl = {https://ui.adsabs.harvard.edu/abs/2025A&A...694A.158D}
}

@ARTICLE{Crestani2021,
       author = {{Crestani}, J. and {Braga}, V.~F. and {Fabrizio}, M. and {Bono}, G. and {Sneden}, C. and {Preston}, G. and {Ferraro}, I. and {Iannicola}, G. and {Nonino}, M. and {Fiorentino}, G. and {Th{\'e}venin}, F. and {Lemasle}, B. and {Prudil}, Z. and {Alves-Brito}, A. and {Altavilla}, G. and {Chaboyer}, B. and {Dall'Ora}, M. and {D'Orazi}, V. and {Gilligan}, C. and {Grebel}, E.~K. and {Koch-Hansen}, A.~J. and {Lala}, H. and {Marengo}, M. and {Marinoni}, S. and {Marrese}, P.~M. and {Mart{\'\i}nez-V{\'a}zquez}, C. and {Matsunaga}, N. and {Monelli}, M. and {Mullen}, J.~P. and {Neeley}, J. and {da Silva}, R. and {Stetson}, P.~B. and {Salaris}, M. and {Storm}, J. and {Valenti}, E. and {Zoccali}, M.},
        title = "{On the Use of Field RR Lyrae as Galactic Probes. III. The {\ensuremath{\alpha}}-element Abundances}",
      journal = {\apj},
         year = 2021,
        month = jun,
       volume = {914},
       number = {1},
          eid = {10},
        pages = {10},
          doi = {10.3847/1538-4357/abfa23},
archivePrefix = {arXiv},
       eprint = {2104.08113},
 primaryClass = {astro-ph.GA},
       adsurl = {https://ui.adsabs.harvard.edu/abs/2021ApJ...914...10C}
}

@ARTICLE{Prudil2020,
       author = {{Prudil}, Z. and {D{\'e}k{\'a}ny}, I. and {Grebel}, E.~K. and {Kunder}, A.},
        title = "{Evidence for Galactic disc RR Lyrae stars in the solar neighbourhood}",
      journal = {\mnras},
         year = 2020,
        month = mar,
       volume = {492},
       number = {3},
        pages = {3408-3419},
          doi = {10.1093/mnras/staa046},
archivePrefix = {arXiv},
       eprint = {2001.02486},
 primaryClass = {astro-ph.SR},
       adsurl = {https://ui.adsabs.harvard.edu/abs/2020MNRAS.492.3408P}
}

@ARTICLE{Haywood2016,
       author = {{Haywood}, M. and {Lehnert}, M.~D. and {Di Matteo}, P. and {Snaith}, O. and {Schultheis}, M. and {Katz}, D. and {G{\'o}mez}, A.},
        title = "{When the Milky Way turned off the lights: APOGEE provides evidence of star formation quenching in our Galaxy}",
      journal = {\aap},
         year = 2016,
        month = may,
       volume = {589},
          eid = {A66},
        pages = {A66},
          doi = {10.1051/0004-6361/201527567},
archivePrefix = {arXiv},
       eprint = {1601.03042},
 primaryClass = {astro-ph.GA},
       adsurl = {https://ui.adsabs.harvard.edu/abs/2016A&A...589A..66H}
}

@ARTICLE{Emma2025,
       author = {{Fern{\'a}ndez-Alvar}, Emma and {Ruiz-Lara}, Tom{\'a}s and {Gallart}, Carme and {Cassisi}, Santi and {Surot}, Francisco and {Gonz{\'a}lez-Koda}, Yllari K. and {Callingham}, Thomas M. and {Queiroz}, Anna B. and {Battaglia}, Giuseppina and {Thomas}, Guillaume and {Chiappini}, Cristina and {Hill}, Vanessa and {Dodd}, Emma and {Helmi}, Amina and {Aznar-Menargues}, Guillem and {de la Cueva}, Alejandro and {Mirabal}, David and {Quintana-Ansaldo}, M{\'o}nica and {Rivero}, Alicia},
        title = "{Chronology of our Galaxy from Gaia colour─magnitude diagram fitting (ChronoGal): II. Unveiling the formation and evolution of the kinematically selected thick and thin discs}",
      journal = {\aap},
         year = 2025,
        month = dec,
       volume = {704},
          eid = {A258},
        pages = {A258},
          doi = {10.1051/0004-6361/202553814},
archivePrefix = {arXiv},
       eprint = {2503.19536},
 primaryClass = {astro-ph.GA},
       adsurl = {https://ui.adsabs.harvard.edu/abs/2025A&A...704A.258F}
}

@ARTICLE{Gunn2006,
       author = {{Gunn}, James E. and {Siegmund}, Walter A. and {Mannery}, Edward J. and {Owen}, Russell E. and {Hull}, Charles L. and {Leger}, R. French and {Carey}, Larry N. and {Knapp}, Gillian R. and {York}, Donald G. and {Boroski}, William N. and {Kent}, Stephen M. and {Lupton}, Robert H. and {Rockosi}, Constance M. and {Evans}, Michael L. and {Waddell}, Patrick and {Anderson}, John E. and {Annis}, James and {Barentine}, John C. and {Bartoszek}, Larry M. and {Bastian}, Steven and {Bracker}, Stephen B. and {Brewington}, Howard J. and {Briegel}, Charles I. and {Brinkmann}, Jon and {Brown}, Yorke J. and {Carr}, Michael A. and {Czarapata}, Paul C. and {Drennan}, Craig C. and {Dombeck}, Thomas and {Federwitz}, Glenn R. and {Gillespie}, Bruce A. and {Gonzales}, Carlos and {Hansen}, Sten U. and {Harvanek}, Michael and {Hayes}, Jeffrey and {Jordan}, Wendell and {Kinney}, Ellyne and {Klaene}, Mark and {Kleinman}, S.~J. and {Kron}, Richard G. and {Kresinski}, Jurek and {Lee}, Glenn and {Limmongkol}, Siriluk and {Lindenmeyer}, Carl W. and {Long}, Daniel C. and {Loomis}, Craig L. and {McGehee}, Peregrine M. and {Mantsch}, Paul M. and {Neilsen}, Jr., Eric H. and {Neswold}, Richard M. and {Newman}, Peter R. and {Nitta}, Atsuko and {Peoples}, Jr., John and {Pier}, Jeffrey R. and {Prieto}, Peter S. and {Prosapio}, Angela and {Rivetta}, Claudio and {Schneider}, Donald P. and {Snedden}, Stephanie and {Wang}, Shu-i.},
        title = "{The 2.5 m Telescope of the Sloan Digital Sky Survey}",
      journal = {\aj},
         year = 2006,
        month = apr,
       volume = {131},
       number = {4},
        pages = {2332-2359},
          doi = {10.1086/500975},
archivePrefix = {arXiv},
       eprint = {astro-ph/0602326},
 primaryClass = {astro-ph},
       adsurl = {https://ui.adsabs.harvard.edu/abs/2006AJ....131.2332G}
}

@ARTICLE{Kroupa1993,
       author = {{Kroupa}, Pavel and {Tout}, Christopher A. and {Gilmore}, Gerard},
        title = "{The Distribution of Low-Mass Stars in the Galactic Disc}",
      journal = {\mnras},
         year = 1993,
        month = jun,
       volume = {262},
        pages = {545-587},
          doi = {10.1093/mnras/262.3.545},
       adsurl = {https://ui.adsabs.harvard.edu/abs/1993MNRAS.262..545K}
}

@ARTICLE{Calura2009,
       author = {{Calura}, F. and {Menci}, N.},
        title = "{Chemical evolution of local galaxies in a hierarchical model}",
      journal = {\mnras},
         year = 2009,
        month = dec,
       volume = {400},
       number = {3},
        pages = {1347-1365},
          doi = {10.1111/j.1365-2966.2009.15440.x},
archivePrefix = {arXiv},
       eprint = {0907.3729},
 primaryClass = {astro-ph.CO},
       adsurl = {https://ui.adsabs.harvard.edu/abs/2009MNRAS.400.1347C}
}

@ARTICLE{Clarke2019,
       author = {{Clarke}, Adam J. and {Debattista}, Victor P. and {Nidever}, David L. and {Loebman}, Sarah R. and {Simons}, Raymond C. and {Kassin}, Susan and {Du}, Min and {Ness}, Melissa and {Fisher}, Deanne B. and {Quinn}, Thomas R. and {Wadsley}, James and {Freeman}, Ken C. and {Popescu}, Cristina C.},
        title = "{The imprint of clump formation at high redshift - I. A disc {\ensuremath{\alpha}}-abundance dichotomy}",
      journal = {\mnras},
         year = 2019,
        month = apr,
       volume = {484},
       number = {3},
        pages = {3476-3490},
          doi = {10.1093/mnras/stz104},
archivePrefix = {arXiv},
       eprint = {1901.00931},
 primaryClass = {astro-ph.GA},
       adsurl = {https://ui.adsabs.harvard.edu/abs/2019MNRAS.484.3476C}
}

@ARTICLE{Johnson2025,
       author = {{Johnson}, James W. and {Feuillet}, Diane K. and {Bonaca}, Ana and {de Brito Silva}, Danielle},
        title = "{That's so Retro: The Gaia-Sausage-Enceladus Merger Trajectory as the Origin of the Chemical Abundance Bimodality in the Milky Way Disk}",
      journal = {arXiv e-prints},
         year = 2025,
        month = oct,
          eid = {arXiv:2510.08688},
        pages = {arXiv:2510.08688},
          doi = {10.48550/arXiv.2510.08688},
archivePrefix = {arXiv},
       eprint = {2510.08688},
 primaryClass = {astro-ph.GA},
       adsurl = {https://ui.adsabs.harvard.edu/abs/2025arXiv251008688J}
}

@ARTICLE{Kennicutt1998,
       author = {{Kennicutt}, Jr., Robert C.},
        title = "{The Global Schmidt Law in Star-forming Galaxies}",
      journal = {\apj},
         year = 1998,
        month = may,
       volume = {498},
       number = {2},
        pages = {541-552},
          doi = {10.1086/305588},
archivePrefix = {arXiv},
       eprint = {astro-ph/9712213},
 primaryClass = {astro-ph},
       adsurl = {https://ui.adsabs.harvard.edu/abs/1998ApJ...498..541K}
}

@ARTICLE{Adibekyan2012,
       author = {{Adibekyan}, V. Zh. and {Sousa}, S.~G. and {Santos}, N.~C. and {Delgado Mena}, E. and {Gonz{\'a}lez Hern{\'a}ndez}, J.~I. and {Israelian}, G. and {Mayor}, M. and {Khachatryan}, G.},
        title = "{Chemical abundances of 1111 FGK stars from the HARPS GTO planet search program. Galactic stellar populations and planets}",
      journal = {\aap},
         year = 2012,
        month = sep,
       volume = {545},
          eid = {A32},
        pages = {A32},
          doi = {10.1051/0004-6361/201219401},
archivePrefix = {arXiv},
       eprint = {1207.2388},
 primaryClass = {astro-ph.EP},
       adsurl = {https://ui.adsabs.harvard.edu/abs/2012A&A...545A..32A}
}

@ARTICLE{SilvaAguirre2018,
       author = {{Silva Aguirre}, V. and {Bojsen-Hansen}, M. and {Slumstrup}, D. and {Casagrande}, L. and {Kawata}, D. and {Ciuc{\v{a}}}, I. and {Handberg}, R. and {Lund}, M.~N. and {Mosumgaard}, J.~R. and {Huber}, D. and {Johnson}, J.~A. and {Pinsonneault}, M.~H. and {Serenelli}, A.~M. and {Stello}, D. and {Tayar}, J. and {Bird}, J.~C. and {Cassisi}, S. and {Hon}, M. and {Martig}, M. and {Nissen}, P.~E. and {Rix}, H.~W. and {Sch{\"o}nrich}, R. and {Sahlholdt}, C. and {Trick}, W.~H. and {Yu}, J.},
        title = "{Confirming chemical clocks: asteroseismic age dissection of the Milky Way disc(s)}",
      journal = {\mnras},
         year = 2018,
        month = apr,
       volume = {475},
       number = {4},
        pages = {5487-5500},
          doi = {10.1093/mnras/sty150},
archivePrefix = {arXiv},
       eprint = {1710.09847},
 primaryClass = {astro-ph.GA},
       adsurl = {https://ui.adsabs.harvard.edu/abs/2018MNRAS.475.5487S}
}

@ARTICLE{Kobayashi2020,
       author = {{Kobayashi}, Chiaki and {Karakas}, Amanda I. and {Lugaro}, Maria},
        title = "{The Origin of Elements from Carbon to Uranium}",
      journal = {\apj},
         year = 2020,
        month = sep,
       volume = {900},
       number = {2},
          eid = {179},
        pages = {179},
          doi = {10.3847/1538-4357/abae65},
archivePrefix = {arXiv},
       eprint = {2008.04660},
 primaryClass = {astro-ph.GA},
       adsurl = {https://ui.adsabs.harvard.edu/abs/2020ApJ...900..179K}
}

@ARTICLE{Ferrini1992,
       author = {{Ferrini}, Federico and {Matteucci}, Francesca and {Pardi}, Chiara and {Penco}, Umberto},
        title = "{Evolution of Spiral Galaxies. I. Halo-Disk Connection for the Evolution of the Solar Neighborhood}",
      journal = {\apj},
         year = 1992,
        month = mar,
       volume = {387},
        pages = {138},
          doi = {10.1086/171066},
       adsurl = {https://ui.adsabs.harvard.edu/abs/1992ApJ...387..138F}
}

@ARTICLE{Pardi1995,
       author = {{Pardi}, Maria Chiara and {Ferrini}, Federico and {Matteucci}, Francesca},
        title = "{Evolution of Spiral Galaxies. IV. The Thick Disk in the Solar Region as an Intermediate Collapse Phase}",
      journal = {\apj},
         year = 1995,
        month = may,
       volume = {444},
        pages = {207},
          doi = {10.1086/175596},
       adsurl = {https://ui.adsabs.harvard.edu/abs/1995ApJ...444..207P}
}

@ARTICLE{Minchev2013,
       author = {{Minchev}, I. and {Chiappini}, C. and {Martig}, M.},
        title = "{Chemodynamical evolution of the Milky Way disk. I. The solar vicinity}",
      journal = {\aap},
         year = 2013,
        month = oct,
       volume = {558},
          eid = {A9},
        pages = {A9},
          doi = {10.1051/0004-6361/201220189},
archivePrefix = {arXiv},
       eprint = {1208.1506},
 primaryClass = {astro-ph.GA},
       adsurl = {https://ui.adsabs.harvard.edu/abs/2013A&A...558A...9M}
}

@ARTICLE{Prantzos2023,
       author = {{Prantzos}, Nikos and {Abia}, Carlos and {Chen}, Tianxiang and {de Laverny}, Patrick and {Recio-Blanco}, Alejandra and {Athanassoula}, E. and {Roberti}, Lorenzo and {Vescovi}, Diego and {Limongi}, Marco and {Chieffi}, Alessandro and {Cristallo}, Sergio},
        title = "{On the origin of the Galactic thin and thick discs, their abundance gradients and the diagnostic potential of their abundance ratios}",
      journal = {\mnras},
         year = 2023,
        month = aug,
       volume = {523},
       number = {2},
        pages = {2126-2145},
          doi = {10.1093/mnras/stad1551},
archivePrefix = {arXiv},
       eprint = {2305.13431},
 primaryClass = {astro-ph.GA},
       adsurl = {https://ui.adsabs.harvard.edu/abs/2023MNRAS.523.2126P}
}

@INPROCEEDINGS{Grisoni2024_proceeding,
       author = {{Grisoni}, Valeria},
        title = "{Galactic archaeology with light and heavy elements}",
    booktitle = {Memorie della Societa Astronomica Italiana},
         year = 2024,
       volume = {95},
        month = mar,
        pages = {41},
          doi = {10.36116/MEMSAIT_95N1.2024.41},
       adsurl = {https://ui.adsabs.harvard.edu/abs/2024MmSAI..95a..41G}
}

@ARTICLE{Roberts2025,
       author = {{Roberts}, John D. and {Pinsonneault}, Marc H. and {Johnson}, Jennifer A. and {Dubay}, Liam O. and {Johnson}, James W.},
        title = "{[C/N] Ages for Red Giants and their Implications for Galactic Archaeology}",
      journal = {arXiv e-prints},
         year = 2025,
        month = sep,
          eid = {arXiv:2509.25321},
        pages = {arXiv:2509.25321},
          doi = {10.48550/arXiv.2509.25321},
archivePrefix = {arXiv},
       eprint = {2509.25321},
 primaryClass = {astro-ph.SR},
       adsurl = {https://ui.adsabs.harvard.edu/abs/2025arXiv250925321R}
}

@ARTICLE{Sharma2021,
       author = {{Sharma}, Sanjib and {Hayden}, Michael R. and {Bland-Hawthorn}, Joss},
        title = "{Chemical enrichment and radial migration in the Galactic disc - the origin of the [{\ensuremath{\alpha}}Fe] double sequence}",
      journal = {\mnras},
         year = 2021,
        month = nov,
       volume = {507},
       number = {4},
        pages = {5882-5901},
          doi = {10.1093/mnras/stab2015},
archivePrefix = {arXiv},
       eprint = {2005.03646},
 primaryClass = {astro-ph.GA},
       adsurl = {https://ui.adsabs.harvard.edu/abs/2021MNRAS.507.5882S}
}

@ARTICLE{Renaud2021,
       author = {{Renaud}, Florent and {Agertz}, Oscar and {Read}, Justin I. and {Ryde}, Nils and {Andersson}, Eric P. and {Bensby}, Thomas and {Rey}, Martin P. and {Feuillet}, Diane K.},
        title = "{VINTERGATAN - II. The history of the Milky Way told by its mergers}",
      journal = {\mnras},
         year = 2021,
        month = jun,
       volume = {503},
       number = {4},
        pages = {5846-5867},
          doi = {10.1093/mnras/stab250},
archivePrefix = {arXiv},
       eprint = {2006.06011},
 primaryClass = {astro-ph.GA},
       adsurl = {https://ui.adsabs.harvard.edu/abs/2021MNRAS.503.5846R}
}

@ARTICLE{Agertz2021,
       author = {{Agertz}, Oscar and {Renaud}, Florent and {Feltzing}, Sofia and {Read}, Justin I. and {Ryde}, Nils and {Andersson}, Eric P. and {Rey}, Martin P. and {Bensby}, Thomas and {Feuillet}, Diane K.},
        title = "{VINTERGATAN - I. The origins of chemically, kinematically, and structurally distinct discs in a simulated Milky Way-mass galaxy}",
      journal = {\mnras},
         year = 2021,
        month = jun,
       volume = {503},
       number = {4},
        pages = {5826-5845},
          doi = {10.1093/mnras/stab322},
archivePrefix = {arXiv},
       eprint = {2006.06008},
 primaryClass = {astro-ph.GA},
       adsurl = {https://ui.adsabs.harvard.edu/abs/2021MNRAS.503.5826A}
}

@ARTICLE{Goswami2021,
       author = {{Goswami}, S. and {Slemer}, A. and {Marigo}, P. and {Bressan}, A. and {Silva}, L. and {Spera}, M. and {Boco}, L. and {Grisoni}, V. and {Pantoni}, L. and {Lapi}, A.},
        title = "{The effects of the initial mass function on Galactic chemical enrichment}",
      journal = {\aap},
         year = 2021,
        month = jun,
       volume = {650},
          eid = {A203},
        pages = {A203},
          doi = {10.1051/0004-6361/202039842},
archivePrefix = {arXiv},
       eprint = {2104.05680},
 primaryClass = {astro-ph.GA},
       adsurl = {https://ui.adsabs.harvard.edu/abs/2021A&A...650A.203G}
}

@ARTICLE{Schoenrich2009,
       author = {{Sch{\"o}nrich}, Ralph and {Binney}, James},
        title = "{Chemical evolution with radial mixing}",
      journal = {\mnras},
         year = 2009,
        month = jun,
       volume = {396},
       number = {1},
        pages = {203-222},
          doi = {10.1111/j.1365-2966.2009.14750.x},
archivePrefix = {arXiv},
       eprint = {0809.3006},
 primaryClass = {astro-ph},
       adsurl = {https://ui.adsabs.harvard.edu/abs/2009MNRAS.396..203S}
}

@ARTICLE{Mishenina2006,
       author = {{Nykytyuk}, T.~V. and {Mishenina}, T.~V.},
        title = "{The Galactic thick and thin disks: differences in evolution}",
      journal = {\aap},
         year = 2006,
        month = sep,
       volume = {456},
       number = {3},
        pages = {969-976},
          doi = {10.1051/0004-6361:20053101},
archivePrefix = {arXiv},
       eprint = {astro-ph/0605661},
 primaryClass = {astro-ph},
       adsurl = {https://ui.adsabs.harvard.edu/abs/2006A&A...456..969N}
}

@ARTICLE{Gent2024,
       author = {{Gent}, Matthew Raymond and {Eitner}, Philipp and {Serenelli}, Aldo and {Friske}, Jennifer K.~S. and {Koposov}, Sergey E. and {Laporte}, Chervin F.~P. and {Buck}, Tobias and {Bergemann}, Maria},
        title = "{The Prince and the Pauper: Evidence for the early high-redshift formation of the Galactic {\ensuremath{\alpha}}-poor disc population}",
      journal = {\aap},
         year = 2024,
        month = mar,
       volume = {683},
          eid = {A74},
        pages = {A74},
          doi = {10.1051/0004-6361/202244157},
archivePrefix = {arXiv},
       eprint = {2206.10949},
 primaryClass = {astro-ph.GA},
       adsurl = {https://ui.adsabs.harvard.edu/abs/2024A&A...683A..74G}
}

@ARTICLE{Beraldo2021,
       author = {{Beraldo e Silva}, Leandro and {Debattista}, Victor P. and {Nidever}, David and {Amarante}, Jo{\~a}o A.~S. and {Garver}, Bethany},
        title = "{Co-formation of the thin and thick discs revealed by APOGEE-DR16 and Gaia-DR2}",
      journal = {\mnras},
         year = 2021,
        month = mar,
       volume = {502},
       number = {1},
        pages = {260-272},
          doi = {10.1093/mnras/staa3966},
archivePrefix = {arXiv},
       eprint = {2009.03346},
 primaryClass = {astro-ph.GA},
       adsurl = {https://ui.adsabs.harvard.edu/abs/2021MNRAS.502..260B}
}

@ARTICLE{Wu2023,
       author = {{Wu}, Yaqian and {Xiang}, Maosheng and {Zhao}, Gang and {Chen}, Yuqin and {Bi}, Shaolan and {Li}, Yaguang},
        title = "{Timing the formation of the galactic thin disc with asteroseismic stellar ages}",
      journal = {\mnras},
         year = 2023,
        month = apr,
       volume = {520},
       number = {2},
        pages = {1913-1927},
          doi = {10.1093/mnras/stad223},
archivePrefix = {arXiv},
       eprint = {2301.07990},
 primaryClass = {astro-ph.GA},
       adsurl = {https://ui.adsabs.harvard.edu/abs/2023MNRAS.520.1913W}
}

@ARTICLE{Matteucci1989,
       author = {{Matteucci}, F. and {Francois}, P.},
        title = "{Galactic chemical evolution : abundance gradients of individual elements.}",
      journal = {\mnras},
         year = 1989,
        month = aug,
       volume = {239},
        pages = {885-904},
          doi = {10.1093/mnras/239.3.885},
       adsurl = {https://ui.adsabs.harvard.edu/abs/1989MNRAS.239..885M}
}

@ARTICLE{Vincenzo2021,
       author = {{Vincenzo}, Fiorenzo and {Weinberg}, David H. and {Miglio}, Andrea and {Lane}, Richard R. and {Roman-Lopes}, Alexandre},
        title = "{The distribution of [{\ensuremath{\alpha}}/Fe] in the Milky Way disc}",
      journal = {\mnras},
         year = 2021,
        month = dec,
       volume = {508},
       number = {4},
        pages = {5903-5920},
          doi = {10.1093/mnras/stab2899},
archivePrefix = {arXiv},
       eprint = {2101.04488},
 primaryClass = {astro-ph.GA},
       adsurl = {https://ui.adsabs.harvard.edu/abs/2021MNRAS.508.5903V}
}

@ARTICLE{Spitoni2023,
       author = {{Spitoni}, E. and {Recio-Blanco}, A. and {de Laverny}, P. and {Palicio}, P.~A. and {Kordopatis}, G. and {Schultheis}, M. and {Contursi}, G. and {Poggio}, E. and {Romano}, D. and {Matteucci}, F.},
        title = "{Beyond the two-infall model. I. Indications for a recent gas infall with Gaia DR3 chemical abundances}",
      journal = {\aap},
         year = 2023,
        month = feb,
       volume = {670},
          eid = {A109},
        pages = {A109},
          doi = {10.1051/0004-6361/202244349},
archivePrefix = {arXiv},
       eprint = {2206.12436},
 primaryClass = {astro-ph.GA},
       adsurl = {https://ui.adsabs.harvard.edu/abs/2023A&A...670A.109S}
}

@ARTICLE{Micali2013,
       author = {{Micali}, A. and {Matteucci}, F. and {Romano}, D.},
        title = "{The chemical evolution of the Milky Way: the Three Infall Model}",
      journal = {\mnras},
         year = 2013,
        month = dec,
       volume = {436},
       number = {2},
        pages = {1648-1658},
          doi = {10.1093/mnras/stt1681},
archivePrefix = {arXiv},
       eprint = {1309.1283},
 primaryClass = {astro-ph.GA},
       adsurl = {https://ui.adsabs.harvard.edu/abs/2013MNRAS.436.1648M}
}

@ARTICLE{Matteucci1986,
       author = {{Matteucci}, F. and {Greggio}, L.},
        title = "{Relative roles of type I and II supernovae in the chemical enrichment of the interstellar gas}",
      journal = {\aap},
         year = 1986,
        month = jan,
       volume = {154},
       number = {1-2},
        pages = {279-287},
       adsurl = {https://ui.adsabs.harvard.edu/abs/1986A&A...154..279M}
}

@ARTICLE{Spitoni2025,
       author = {{Spitoni}, E. and {Palla}, M. and {Magrini}, L. and {Matteucci}, F. and {Danielski}, C. and {Tsantaki}, M. and {Sozzetti}, A. and {Molero}, M. and {Fontani}, F. and {Romano}, D. and {Cescutti}, G. and {Silva}, L.},
        title = "{Shaping Galactic habitability: Impact of stellar migration and gas giants}",
      journal = {\aap},
         year = 2025,
        month = aug,
       volume = {700},
          eid = {A58},
        pages = {A58},
          doi = {10.1051/0004-6361/202555050},
archivePrefix = {arXiv},
       eprint = {2506.19981},
 primaryClass = {astro-ph.GA},
       adsurl = {https://ui.adsabs.harvard.edu/abs/2025A&A...700A..58S}
}

@ARTICLE{Soderblom2010,
       author = {{Soderblom}, David R.},
        title = "{The Ages of Stars}",
      journal = {\araa},
         year = 2010,
        month = sep,
       volume = {48},
        pages = {581-629},
          doi = {10.1146/annurev-astro-081309-130806},
archivePrefix = {arXiv},
       eprint = {1003.6074},
 primaryClass = {astro-ph.SR},
       adsurl = {https://ui.adsabs.harvard.edu/abs/2010ARA&A..48..581S}
}

@ARTICLE{Prantzos2025,
       author = {{Chen}, Tianxiang and {Prantzos}, Nikos},
        title = "{Recent star formation episodes in the Galaxy: Impact on its chemical properties and the evolution of its abundance gradient}",
      journal = {\aap},
         year = 2025,
        month = feb,
       volume = {694},
          eid = {A120},
        pages = {A120},
          doi = {10.1051/0004-6361/202452552},
archivePrefix = {arXiv},
       eprint = {2501.03342},
 primaryClass = {astro-ph.GA},
       adsurl = {https://ui.adsabs.harvard.edu/abs/2025A&A...694A.120C}
}

@ARTICLE{Gilmore1983,
       author = {{Gilmore}, G. and {Reid}, N.},
        title = "{New light on faint stars - III. Galactic structure towards the South Pole and the Galactic thick disc.}",
      journal = {\mnras},
         year = 1983,
        month = mar,
       volume = {202},
        pages = {1025-1047},
          doi = {10.1093/mnras/202.4.1025},
       adsurl = {https://ui.adsabs.harvard.edu/abs/1983MNRAS.202.1025G}
}

@ARTICLE{Palla2020,
       author = {{Palla}, M. and {Matteucci}, F. and {Spitoni}, E. and {Vincenzo}, F. and {Grisoni}, V.},
        title = "{Chemical evolution of the Milky Way: constraints on the formation of the thick and thin discs}",
      journal = {\mnras},
         year = 2020,
        month = oct,
       volume = {498},
       number = {2},
        pages = {1710-1725},
          doi = {10.1093/mnras/staa2437},
archivePrefix = {arXiv},
       eprint = {2008.07484},
 primaryClass = {astro-ph.GA},
       adsurl = {https://ui.adsabs.harvard.edu/abs/2020MNRAS.498.1710P}
}

@ARTICLE{Grisoni2024,
       author = {{Grisoni}, V. and {Chiappini}, C. and {Miglio}, A. and {Brogaard}, K. and {Casali}, G. and {Willett}, E. and {Montalb{\'a}n}, J. and {Stokholm}, A. and {Thomsen}, J.~S. and {Tailo}, M. and {Matteuzzi}, M. and {Valentini}, M. and {Elsworth}, Y. and {Mosser}, B.},
        title = "{K2 results for ``young'' {\ensuremath{\alpha}}-rich stars in the Galaxy}",
      journal = {\aap},
         year = 2024,
        month = mar,
       volume = {683},
          eid = {A111},
        pages = {A111},
          doi = {10.1051/0004-6361/202347440},
archivePrefix = {arXiv},
       eprint = {2312.07091},
 primaryClass = {astro-ph.GA},
       adsurl = {https://ui.adsabs.harvard.edu/abs/2024A&A...683A.111G}
}

@ARTICLE{Willett2023,
       author = {{Willett}, Emma and {Miglio}, Andrea and {Mackereth}, J. Ted and {Chiappini}, Cristina and {Lyttle}, Alexander J. and {Elsworth}, Yvonne and {Mosser}, Beno{\^\i}t and {Khan}, Saniya and {Anders}, Friedrich and {Casali}, Giada and {Grisoni}, Valeria},
        title = "{The evolution of the Milky Way's thin disc radial metallicity gradient with K2 asteroseismic ages}",
      journal = {\mnras},
         year = 2023,
        month = dec,
       volume = {526},
       number = {2},
        pages = {2141-2155},
          doi = {10.1093/mnras/stad2374},
archivePrefix = {arXiv},
       eprint = {2307.14422},
 primaryClass = {astro-ph.GA},
       adsurl = {https://ui.adsabs.harvard.edu/abs/2023MNRAS.526.2141W}
}

@ARTICLE{Ruiz-Lara2020,
       author = {{Ruiz-Lara}, Tom{\'a}s and {Gallart}, Carme and {Bernard}, Edouard J. and {Cassisi}, Santi},
        title = "{The recurrent impact of the Sagittarius dwarf on the star formation history of the Milky Way}",
      journal = {Nature Astronomy},
         year = 2020,
        month = may,
       volume = {4},
        pages = {965-973},
          doi = {10.1038/s41550-020-1097-0},
archivePrefix = {arXiv},
       eprint = {2003.12577},
 primaryClass = {astro-ph.GA},
       adsurl = {https://ui.adsabs.harvard.edu/abs/2020NatAs...4..965R}
}

@ARTICLE{Helmi2018,
       author = {{Helmi}, Amina and {Babusiaux}, Carine and {Koppelman}, Helmer H. and {Massari}, Davide and {Veljanoski}, Jovan and {Brown}, Anthony G.~A.},
        title = "{The merger that led to the formation of the Milky Way's inner stellar halo and thick disk}",
      journal = {\nat},
         year = 2018,
        month = oct,
       volume = {563},
       number = {7729},
        pages = {85-88},
          doi = {10.1038/s41586-018-0625-x},
archivePrefix = {arXiv},
       eprint = {1806.06038},
 primaryClass = {astro-ph.GA},
       adsurl = {https://ui.adsabs.harvard.edu/abs/2018Natur.563...85H}
}

@ARTICLE{Vincenzo2019,
       author = {{Vincenzo}, Fiorenzo and {Spitoni}, Emanuele and {Calura}, Francesco and {Matteucci}, Francesca and {Silva Aguirre}, Victor and {Miglio}, Andrea and {Cescutti}, Gabriele},
        title = "{The Fall of a Giant. Chemical evolution of Enceladus, alias the Gaia Sausage}",
      journal = {\mnras},
         year = 2019,
        month = jul,
       volume = {487},
       number = {1},
        pages = {L47-L52},
          doi = {10.1093/mnrasl/slz070},
archivePrefix = {arXiv},
       eprint = {1903.03465},
 primaryClass = {astro-ph.GA},
       adsurl = {https://ui.adsabs.harvard.edu/abs/2019MNRAS.487L..47V}
}

@ARTICLE{Spitoni2024,
       author = {{Spitoni}, E. and {Matteucci}, F. and {Gratton}, R. and {Ratcliffe}, B. and {Minchev}, I. and {Cescutti}, G.},
        title = "{(Re)mind the gap: A hiatus in star formation history unveiled by APOGEE DR17}",
      journal = {\aap},
         year = 2024,
        month = oct,
       volume = {690},
          eid = {A208},
        pages = {A208},
          doi = {10.1051/0004-6361/202450754},
archivePrefix = {arXiv},
       eprint = {2405.11025},
 primaryClass = {astro-ph.GA},
       adsurl = {https://ui.adsabs.harvard.edu/abs/2024A&A...690A.208S}
}

@ARTICLE{Nepal2024,
       author = {{Nepal}, S. and {Chiappini}, C. and {Queiroz}, A.~B. and {Guiglion}, G. and {Montalb{\'a}n}, J. and {Steinmetz}, M. and {Miglio}, A. and {Khalatyan}, A.},
        title = "{Discovery of the local counterpart of disc galaxies at z > 4: The oldest thin disc of the Milky Way using Gaia-RVS}",
      journal = {\aap},
         year = 2024,
        month = aug,
       volume = {688},
          eid = {A167},
        pages = {A167},
          doi = {10.1051/0004-6361/202449445},
archivePrefix = {arXiv},
       eprint = {2402.00561},
 primaryClass = {astro-ph.GA},
       adsurl = {https://ui.adsabs.harvard.edu/abs/2024A&A...688A.167N}
}

@ARTICLE{Zhang2024,
       author = {{Zhang}, Hanyuan and {Ardern-Arentsen}, Anke and {Belokurov}, Vasily},
        title = "{On the existence of a very metal-poor disc in the Milky Way}",
      journal = {\mnras},
         year = 2024,
        month = sep,
       volume = {533},
       number = {1},
        pages = {889-907},
          doi = {10.1093/mnras/stae1887},
archivePrefix = {arXiv},
       eprint = {2311.09294},
 primaryClass = {astro-ph.GA},
       adsurl = {https://ui.adsabs.harvard.edu/abs/2024MNRAS.533..889Z}
}

@ARTICLE{Rizzuti2025,
       author = {{Rizzuti}, F. and {Cescutti}, G. and {Molaro}, P. and {Roberti}, L. and {Chieffi}, A. and {Limongi}, M. and {Magrini}, L. and {Matteucci}, F.},
        title = "{Explaining the $^{12}$C/$^{13}$C ratio in the Galactic halo: The contribution from shell mergers in primordial massive stars}",
      journal = {\aap},
         year = 2025,
        month = jun,
       volume = {698},
          eid = {A118},
        pages = {A118},
          doi = {10.1051/0004-6361/202453603},
archivePrefix = {arXiv},
       eprint = {2503.20876},
 primaryClass = {astro-ph.GA},
       adsurl = {https://ui.adsabs.harvard.edu/abs/2025A&A...698A.118R}
}

@ARTICLE{Grisoni2025,
       author = {{Grisoni}, V. and {Rizzuti}, F. and {Cescutti}, G.},
        title = "{Fluorine evolution in the Galactic halo}",
      journal = {\aap},
         year = 2025,
        month = dec,
       volume = {704},
          eid = {A45},
        pages = {A45},
          doi = {10.1051/0004-6361/202557691},
archivePrefix = {arXiv},
       eprint = {2510.27300},
 primaryClass = {astro-ph.GA},
       adsurl = {https://ui.adsabs.harvard.edu/abs/2025A&A...704A..45G}
}

@ARTICLE{Hayden2017,
       author = {{Hayden}, M.~R. and {Recio-Blanco}, A. and {de Laverny}, P. and {Mikolaitis}, S. and {Worley}, C.~C.},
        title = "{The AMBRE project: The thick thin disk and thin thick disk of the Milky Way}",
      journal = {\aap},
         year = 2017,
        month = dec,
       volume = {608},
          eid = {L1},
        pages = {L1},
          doi = {10.1051/0004-6361/201731494},
archivePrefix = {arXiv},
       eprint = {1712.02358},
 primaryClass = {astro-ph.GA},
       adsurl = {https://ui.adsabs.harvard.edu/abs/2017A&A...608L...1H}
}

@ARTICLE{Bono2026,
       author = {{Bono}, G. and {Braga}, V.~F. and {Fabrizio}, M. and {Tantalo}, M. and {Baeza-Villagra}, K. and {Crestani}, J. and {D'Orazi}, V. and {Dall'Ora}, M. and {Di Criscienzo}, M. and {Fiorentino}, G. and {Gholami}, M. and {Marengo}, M. and {Mart{\'\i}nez-V{\'a}zquez}, C.~E. and {Monelli}, M. and {Mullen}, J.~P. and {Nunnari}, A. and {Pipwala}, V.~D. and {Prudil}, Z. and {Sneden}, C. and {Altavilla}, G. and {Bergemann}, M. and {B{\"o}cek Topcu}, G. and {Buonanno}, R. and {Calamida}, A. and {Carretta}, E. and {Ceci}, G. and {Chaboyer}, B. and {Correnti}, M. and {da Silva}, R. and {Ferraro}, I. and {G{\'o}mez}, F.~A. and {Iannicola}, G. and {Kudritzki}, R.-P. and {Kunder}, A. and {Kwak}, S. and {Marconi}, M. and {Marinoni}, S. and {Matsunaga}, N. and {Matteucci}, F. and {Monachesi}, A. and {Musella}, I. and {Navarro Ovando}, M.~G. and {Preston}, G.~W. and {Ripepi}, V. and {Salaris}, M. and {S{\'a}nchez-Benavente}, M. and {Spitoni}, E. and {Stetson}, P.~B. and {Th{\'e}venin}, F. and {Thompson}, I.~B. and {Tissera}, P.~B. and {Tsujimoto}, T. and {Valenti}, E. and {Vivas}, A.~K. and {Walker}, A.~R. and {Zoccali}, M. and {Zocchi}, A.},
        title = "{On the Use of Field RR Lyrae as Galactic Probes -- VIII. Early Formation of the Galactic Spheroid}",
      journal = {arXiv e-prints},
         year = 2026,
        month = jan,
          eid = {arXiv:2601.16523},
        pages = {arXiv:2601.16523},
          doi = {10.48550/arXiv.2601.16523},
archivePrefix = {arXiv},
       eprint = {2601.16523},
 primaryClass = {astro-ph.GA},
       adsurl = {https://ui.adsabs.harvard.edu/abs/2026arXiv260116523B}
}

@ARTICLE{Arentsen2024,
       author = {{Ardern-Arentsen}, Anke and {Monari}, Giacomo and {Queiroz}, Anna B.~A. and {Starkenburg}, Else and {Martin}, Nicolas F. and {Chiappini}, Cristina and {Aguado}, David S. and {Belokurov}, Vasily and {Carlberg}, Ray and {Monty}, Stephanie and {Myeong}, GyuChul and {Schultheis}, Mathias and {Sestito}, Federico and {Venn}, Kim A. and {Vitali}, Sara and {Yuan}, Zhen and {Zhang}, Hanyuan and {Buder}, Sven and {Lewis}, Geraint F. and {Oliver}, William H. and {Wan}, Zhen and {Zucker}, Daniel B.},
        title = "{The Pristine Inner Galaxy Survey - VIII. Characterizing the orbital properties of the ancient, very metal-poor inner Milky Way}",
      journal = {\mnras},
         year = 2024,
        month = may,
       volume = {530},
       number = {3},
        pages = {3391-3411},
          doi = {10.1093/mnras/stae1049},
archivePrefix = {arXiv},
       eprint = {2312.03847},
 primaryClass = {astro-ph.GA},
       adsurl = {https://ui.adsabs.harvard.edu/abs/2024MNRAS.530.3391A}
}

@ARTICLE{Viswanathan2025,
       author = {{Viswanathan}, Akshara and {Horta}, Danny and {Price-Whelan}, Adrian M. and {Starkenburg}, Else},
        title = "{A slow spin to win: The gradual kinematic evolution across metallicities of the proto-Galaxy to the high-{\ensuremath{\alpha}} disc}",
      journal = {\aap},
         year = 2025,
        month = nov,
       volume = {703},
          eid = {A183},
        pages = {A183},
          doi = {10.1051/0004-6361/202453063},
archivePrefix = {arXiv},
       eprint = {2411.12165},
 primaryClass = {astro-ph.GA},
       adsurl = {https://ui.adsabs.harvard.edu/abs/2025A&A...703A.183V}
}

@ARTICLE{Grisoni2017,
       author = {{Grisoni}, V. and {Spitoni}, E. and {Matteucci}, F. and {Recio-Blanco}, A. and {de Laverny}, P. and {Hayden}, M. and {Mikolaitis}, {\^{S}}. and {Worley}, C.~C.},
        title = "{The AMBRE project: chemical evolution models for the Milky Way thick and thin discs}",
      journal = {\mnras},
         year = 2017,
        month = dec,
       volume = {472},
       number = {3},
        pages = {3637-3647},
          doi = {10.1093/mnras/stx2201},
archivePrefix = {arXiv},
       eprint = {1706.02614},
 primaryClass = {astro-ph.GA},
       adsurl = {https://ui.adsabs.harvard.edu/abs/2017MNRAS.472.3637G}
}

@ARTICLE{Grisoni2018,
       author = {{Grisoni}, V. and {Spitoni}, E. and {Matteucci}, F.},
        title = "{Abundance gradients along the Galactic disc from chemical evolution models}",
      journal = {\mnras},
         year = 2018,
        month = dec,
       volume = {481},
       number = {2},
        pages = {2570-2580},
          doi = {10.1093/mnras/sty2444},
archivePrefix = {arXiv},
       eprint = {1805.11415},
 primaryClass = {astro-ph.GA},
       adsurl = {https://ui.adsabs.harvard.edu/abs/2018MNRAS.481.2570G}
}

@ARTICLE{Spitoni2019,
       author = {{Spitoni}, E. and {Silva Aguirre}, V. and {Matteucci}, F. and {Calura}, F. and {Grisoni}, V.},
        title = "{Galactic Archaeology with asteroseismic ages: Evidence for delayed gas infall in the formation of the Milky Way disc}",
      journal = {\aap},
         year = 2019,
        month = mar,
       volume = {623},
          eid = {A60},
        pages = {A60},
          doi = {10.1051/0004-6361/201834188},
archivePrefix = {arXiv},
       eprint = {1809.00914},
 primaryClass = {astro-ph.GA},
       adsurl = {https://ui.adsabs.harvard.edu/abs/2019A&A...623A..60S}
}

@ARTICLE{Matteucci2019,
       author = {{Matteucci}, F. and {Grisoni}, V. and {Spitoni}, E. and {Zulianello}, A. and {Rojas-Arriagada}, A. and {Schultheis}, M. and {Ryde}, N.},
        title = "{The origin of stellar populations in the Galactic bulge from chemical abundances}",
      journal = {\mnras},
         year = 2019,
        month = aug,
       volume = {487},
       number = {4},
        pages = {5363-5371},
          doi = {10.1093/mnras/stz1647},
archivePrefix = {arXiv},
       eprint = {1903.12440},
 primaryClass = {astro-ph.GA},
       adsurl = {https://ui.adsabs.harvard.edu/abs/2019MNRAS.487.5363M}
}

@ARTICLE{Grisoni2019,
       author = {{Grisoni}, V. and {Matteucci}, F. and {Romano}, D. and {Fu}, X.},
        title = "{Evolution of lithium in the Milky Way halo, discs, and bulge}",
      journal = {\mnras},
         year = 2019,
        month = nov,
       volume = {489},
       number = {3},
        pages = {3539-3546},
          doi = {10.1093/mnras/stz2428},
archivePrefix = {arXiv},
       eprint = {1906.09130},
 primaryClass = {astro-ph.GA},
       adsurl = {https://ui.adsabs.harvard.edu/abs/2019MNRAS.489.3539G}
}

@ARTICLE{Grisoni2020a,
       author = {{Grisoni}, V. and {Cescutti}, G. and {Matteucci}, F. and {Forsberg}, R. and {J{\"o}nsson}, H. and {Ryde}, N.},
        title = "{Modelling the chemical evolution of Zr, La, Ce, and Eu in the Galactic discs and bulge}",
      journal = {\mnras},
         year = 2020,
        month = feb,
       volume = {492},
       number = {2},
        pages = {2828-2834},
          doi = {10.1093/mnras/staa051},
archivePrefix = {arXiv},
       eprint = {1911.04271},
 primaryClass = {astro-ph.GA},
       adsurl = {https://ui.adsabs.harvard.edu/abs/2020MNRAS.492.2828G}
}

@ARTICLE{Grisoni2020b,
       author = {{Grisoni}, V. and {Romano}, D. and {Spitoni}, E. and {Matteucci}, F. and {Ryde}, N. and {J{\"o}nsson}, H.},
        title = "{Fluorine in the solar neighbourhood: modelling the Galactic thick and thin discs}",
      journal = {\mnras},
         year = 2020,
        month = oct,
       volume = {498},
       number = {1},
        pages = {1252-1258},
          doi = {10.1093/mnras/staa2316},
archivePrefix = {arXiv},
       eprint = {2008.00812},
 primaryClass = {astro-ph.GA},
       adsurl = {https://ui.adsabs.harvard.edu/abs/2020MNRAS.498.1252G}
}

@ARTICLE{Grisoni2021,
       author = {{Grisoni}, V. and {Matteucci}, F. and {Romano}, D.},
        title = "{Nitrogen evolution in the halo, thick disc, thin disc, and bulge of the Galaxy}",
      journal = {\mnras},
         year = 2021,
        month = nov,
       volume = {508},
       number = {1},
        pages = {719-727},
          doi = {10.1093/mnras/stab2579},
archivePrefix = {arXiv},
       eprint = {2109.03642},
 primaryClass = {astro-ph.GA},
       adsurl = {https://ui.adsabs.harvard.edu/abs/2021MNRAS.508..719G}
}

@ARTICLE{Spitoni2021,
       author = {{Spitoni}, E. and {Verma}, K. and {Silva Aguirre}, V. and {Vincenzo}, F. and {Matteucci}, F. and {Vai{\v{c}}ekauskait{\.{e}}}, B. and {Palla}, M. and {Grisoni}, V. and {Calura}, F.},
        title = "{APOGEE DR16: A multi-zone chemical evolution model for the Galactic disc based on MCMC methods}",
      journal = {\aap},
         year = 2021,
        month = mar,
       volume = {647},
          eid = {A73},
        pages = {A73},
          doi = {10.1051/0004-6361/202039864},
archivePrefix = {arXiv},
       eprint = {2101.08803},
 primaryClass = {astro-ph.GA},
       adsurl = {https://ui.adsabs.harvard.edu/abs/2021A&A...647A..73S}
}

@ARTICLE{Cerqui2025,
       author = {{Cerqui}, Valeria and {Haywood}, Misha and {Snaith}, Owain and {Di Matteo}, Paola and {Casamiquela}, Laia},
        title = "{The chemical enrichment histories across the Milky Way disk}",
      journal = {\aap},
         year = 2025,
        month = jul,
       volume = {699},
          eid = {A277},
        pages = {A277},
          doi = {10.1051/0004-6361/202452448},
archivePrefix = {arXiv},
       eprint = {2504.20160},
 primaryClass = {astro-ph.GA},
       adsurl = {https://ui.adsabs.harvard.edu/abs/2025A&A...699A.277C}
}

@ARTICLE{Warfield2024,
       author = {{Warfield}, Jack T. and {Zinn}, Joel C. and {Schonhut-Stasik}, Jessica and {Johnson}, James W. and {Pinsonneault}, Marc H. and {Johnson}, Jennifer A. and {Stello}, Dennis and {Beaton}, Rachael L. and {Elsworth}, Yvonne and {Garc{\'\i}a}, Rafael A. and {Mathur}, Savita and {Mosser}, Beno{\^\i}t and {Serenelli}, Aldo and {Tayar}, Jamie},
        title = "{The APO-K2 Catalog. II. Accurate Stellar Ages for Red Giant Branch Stars across the Milky Way}",
      journal = {\aj},
         year = 2024,
        month = may,
       volume = {167},
       number = {5},
          eid = {208},
        pages = {208},
          doi = {10.3847/1538-3881/ad33bb},
archivePrefix = {arXiv},
       eprint = {2403.16250},
 primaryClass = {astro-ph.GA},
       adsurl = {https://ui.adsabs.harvard.edu/abs/2024AJ....167..208W}
}

@ARTICLE{Anders2017a,
       author = {{Anders}, F. and {Chiappini}, C. and {Rodrigues}, T.~S. and {Miglio}, A. and {Montalb{\'a}n}, J. and {Mosser}, B. and {Girardi}, L. and {Valentini}, M. and {Noels}, A. and {Morel}, T. and {Johnson}, J.~A. and {Schultheis}, M. and {Baudin}, F. and {de Assis Peralta}, R. and {Hekker}, S. and {Theme{\ss}l}, N. and {Kallinger}, T. and {Garc{\'\i}a}, R.~A. and {Mathur}, S. and {Baglin}, A. and {Santiago}, B.~X. and {Martig}, M. and {Minchev}, I. and {Steinmetz}, M. and {da Costa}, L.~N. and {Maia}, M.~A.~G. and {Allende Prieto}, C. and {Cunha}, K. and {Beers}, T.~C. and {Epstein}, C. and {Garc{\'\i}a P{\'e}rez}, A.~E. and {Garc{\'\i}a-Hern{\'a}ndez}, D.~A. and {Harding}, P. and {Holtzman}, J. and {Majewski}, S.~R. and {M{\'e}sz{\'a}ros}, Sz. and {Nidever}, D. and {Pan}, K. and {Pinsonneault}, M. and {Schiavon}, R.~P. and {Schneider}, D.~P. and {Shetrone}, M.~D. and {Stassun}, K. and {Zamora}, O. and {Zasowski}, G.},
        title = "{Galactic archaeology with asteroseismology and spectroscopy: Red giants observed by CoRoT and APOGEE}",
      journal = {\aap},
         year = 2017,
        month = jan,
       volume = {597},
          eid = {A30},
        pages = {A30},
          doi = {10.1051/0004-6361/201527204},
archivePrefix = {arXiv},
       eprint = {1604.07763},
 primaryClass = {astro-ph.GA},
       adsurl = {https://ui.adsabs.harvard.edu/abs/2017A&A...597A..30A}
}

@ARTICLE{Anders2017b,
       author = {{Anders}, F. and {Chiappini}, C. and {Minchev}, I. and {Miglio}, A. and {Montalb{\'a}n}, J. and {Mosser}, B. and {Rodrigues}, T.~S. and {Santiago}, B.~X. and {Baudin}, F. and {Beers}, T.~C. and {da Costa}, L.~N. and {Garc{\'\i}a}, R.~A. and {Garc{\'\i}a-Hern{\'a}ndez}, D.~A. and {Holtzman}, J. and {Maia}, M.~A.~G. and {Majewski}, S. and {Mathur}, S. and {Noels-Grotsch}, A. and {Pan}, K. and {Schneider}, D.~P. and {Schultheis}, M. and {Steinmetz}, M. and {Valentini}, M. and {Zamora}, O.},
        title = "{Red giants observed by CoRoT and APOGEE: The evolution of the Milky Way's radial metallicity gradient}",
      journal = {\aap},
         year = 2017,
        month = apr,
       volume = {600},
          eid = {A70},
        pages = {A70},
          doi = {10.1051/0004-6361/201629363},
archivePrefix = {arXiv},
       eprint = {1608.04951},
 primaryClass = {astro-ph.GA},
       adsurl = {https://ui.adsabs.harvard.edu/abs/2017A&A...600A..70A}
}

@ARTICLE{Miglio2021,
       author = {{Miglio}, A. and {Chiappini}, C. and {Mackereth}, J.~T. and {Davies}, G.~R. and {Brogaard}, K. and {Casagrande}, L. and {Chaplin}, W.~J. and {Girardi}, L. and {Kawata}, D. and {Khan}, S. and {Izzard}, R. and {Montalb{\'a}n}, J. and {Mosser}, B. and {Vincenzo}, F. and {Bossini}, D. and {Noels}, A. and {Rodrigues}, T. and {Valentini}, M. and {Mandel}, I.},
        title = "{Age dissection of the Milky Way discs: Red giants in the Kepler field}",
      journal = {\aap},
         year = 2021,
        month = jan,
       volume = {645},
          eid = {A85},
        pages = {A85},
          doi = {10.1051/0004-6361/202038307},
archivePrefix = {arXiv},
       eprint = {2004.14806},
 primaryClass = {astro-ph.GA},
       adsurl = {https://ui.adsabs.harvard.edu/abs/2021A&A...645A..85M}
}

@ARTICLE{Matteucci2021,
       author = {{Matteucci}, Francesca},
        title = "{Modelling the chemical evolution of the Milky Way}",
      journal = {\aapr},
         year = 2021,
        month = dec,
       volume = {29},
       number = {1},
          eid = {5},
        pages = {5},
          doi = {10.1007/s00159-021-00133-8},
archivePrefix = {arXiv},
       eprint = {2106.13145},
 primaryClass = {astro-ph.GA},
       adsurl = {https://ui.adsabs.harvard.edu/abs/2021A&ARv..29....5M}
}

@ARTICLE{Chiappini2015,
       author = {{Chiappini}, C. and {Anders}, F. and {Rodrigues}, T.~S. and {Miglio}, A. and {Montalb{\'a}n}, J. and {Mosser}, B. and {Girardi}, L. and {Valentini}, M. and {Noels}, A. and {Morel}, T. and {Minchev}, I. and {Steinmetz}, M. and {Santiago}, B.~X. and {Schultheis}, M. and {Martig}, M. and {da Costa}, L.~N. and {Maia}, M.~A.~G. and {Allende Prieto}, C. and {de Assis Peralta}, R. and {Hekker}, S. and {Theme{\ss}l}, N. and {Kallinger}, T. and {Garc{\'\i}a}, R.~A. and {Mathur}, S. and {Baudin}, F. and {Beers}, T.~C. and {Cunha}, K. and {Harding}, P. and {Holtzman}, J. and {Majewski}, S. and {M{\'e}sz{\'a}ros}, Sz. and {Nidever}, D. and {Pan}, K. and {Schiavon}, R.~P. and {Shetrone}, M.~D. and {Schneider}, D.~P. and {Stassun}, K.},
        title = "{Young [{\ensuremath{\alpha}}/Fe]-enhanced stars discovered by CoRoT and APOGEE: What is their origin?}",
      journal = {\aap},
         year = 2015,
        month = apr,
       volume = {576},
          eid = {L12},
        pages = {L12},
          doi = {10.1051/0004-6361/201525865},
archivePrefix = {arXiv},
       eprint = {1503.06990},
 primaryClass = {astro-ph.SR},
       adsurl = {https://ui.adsabs.harvard.edu/abs/2015A&A...576L..12C}
}

\begin{appendix}

\section{Model for a massive dwarf galaxy}

\begin{figure}
\centering
    \includegraphics[scale=0.28]{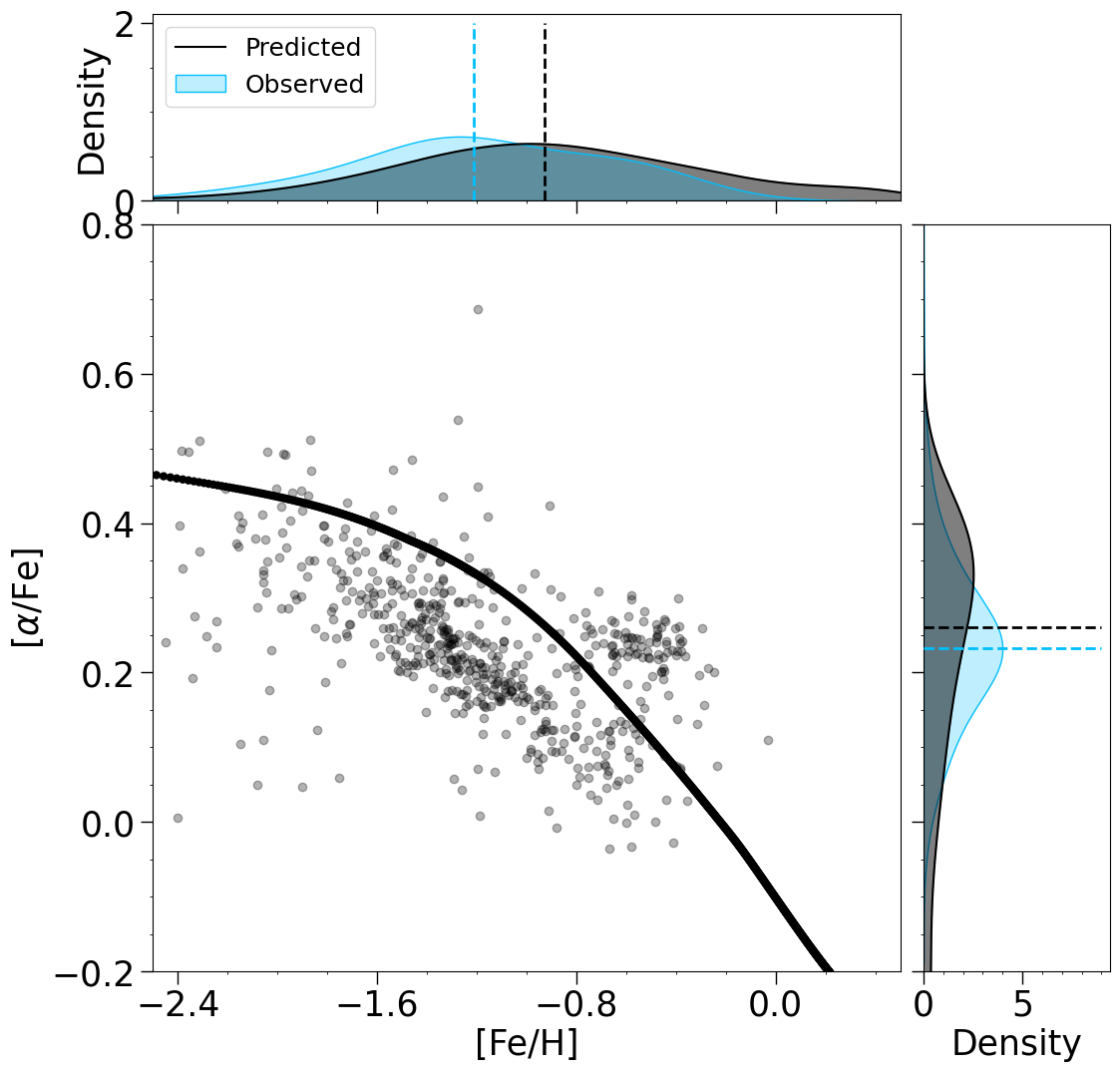}
    \caption{[$\alpha$/Fe] vs. [Fe/H] for the reference model for a massive dwarf compared to data from \cite{Helmi2018}.
    }
    \label{enceladus}
\end{figure}

In our model for the Galactic thick and thin discs, by following the approach described in \cite{Spitoni2024}, we consider that the Galactic disc has been built both from primordial gas and pre-enriched one.
In order to get our pre-enrichment levels, we then run the chemical evolution of a massive dwarf galaxy which is then accreted after a certain evolutionary time and provides some level of pre-enrichment. We consider for such a galaxy a one-infall model with parameters consistent with the ones of \cite{Vincenzo2019}, where they proposed a first model for the chemical evolution of Gaia-Enceladus. This model is characterized by a low star formation efficiency ($\nu$=0.6 Gyr$^{-1}$), short infall timescale ($\tau$=0.24 Gyr) and the presence of a galactic wind (loading factor $\omega$=2), which allows to reproduce the abundance pattern as well as the metallicity distribution function of Gaia-Enceladus stars \citep{Vincenzo2019,Cescutti2020,Spitoni2024}.
In Fig. \ref{enceladus}, we show the [$\alpha$/Fe] vs. [Fe/H] predicted for our massive dwarf galaxy model, where we can then obtain a good agreement between model predictions and observations.

\section{Separate analysis for thick and thin disc models}

\begin{figure*}
\centering
 	\includegraphics[scale=0.3]{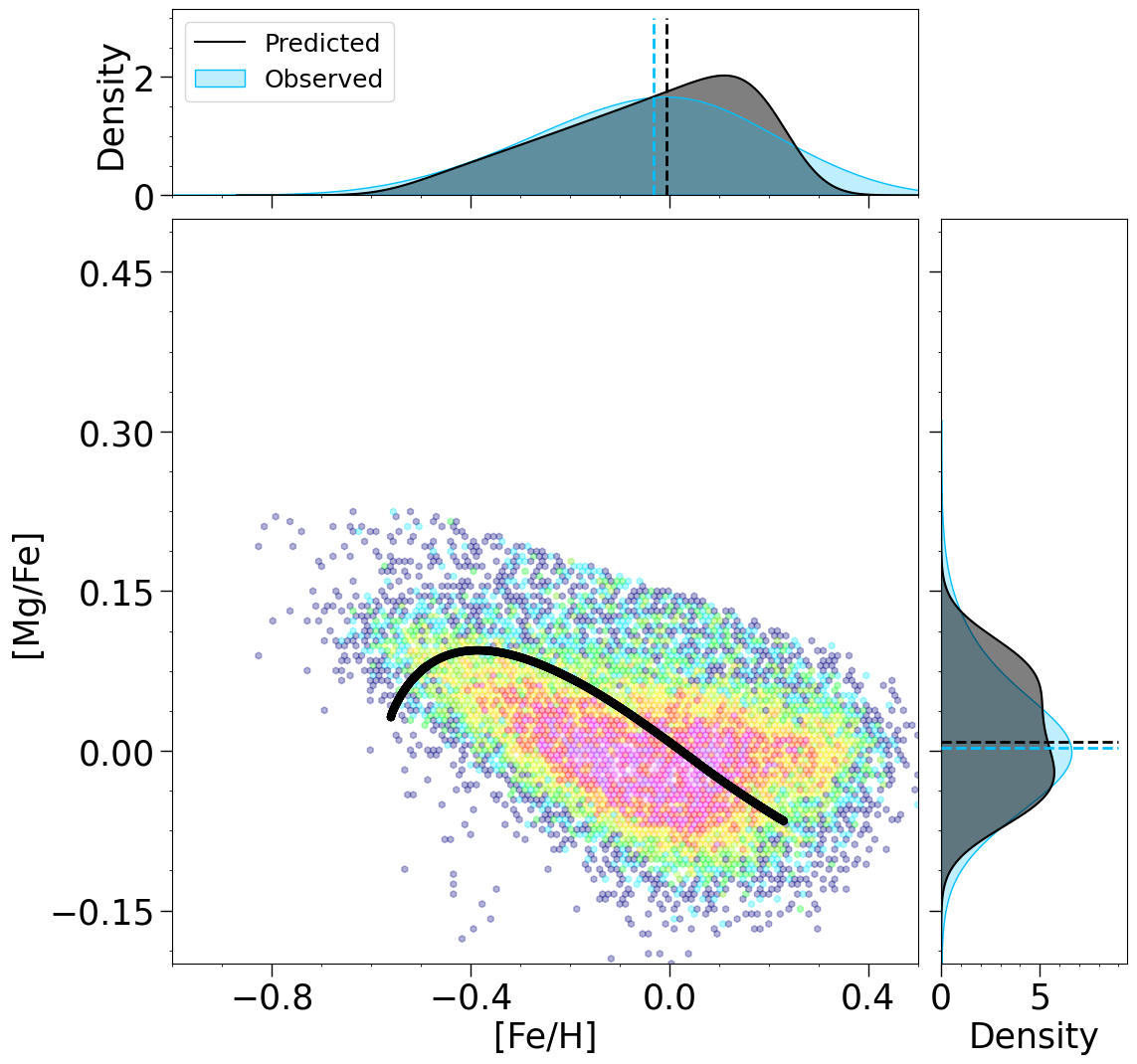}
    \includegraphics[scale=0.3]{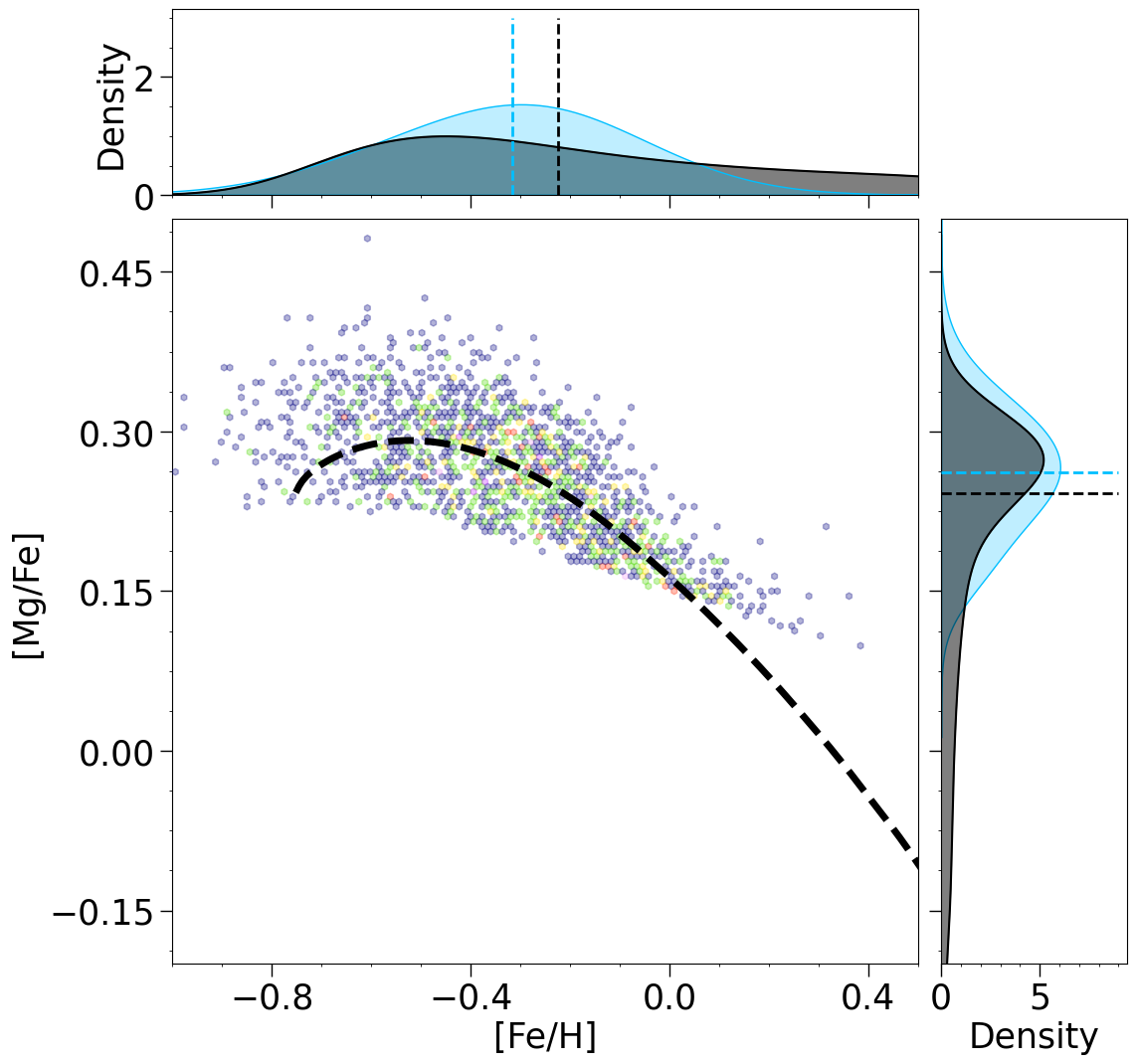}
    \caption{Observed and predicted [$\alpha$/Fe] vs. [Fe/H]  for the low- and high-alpha sequences (left and right panels, respectively). Data are from \cite{Borbolato2025}. The predictions are from the revised parallel model for the Galactic thick and thin discs.
    }
    \label{appendixB}
\end{figure*}

\begin{figure*}
\centering
                \includegraphics[scale=0.3]{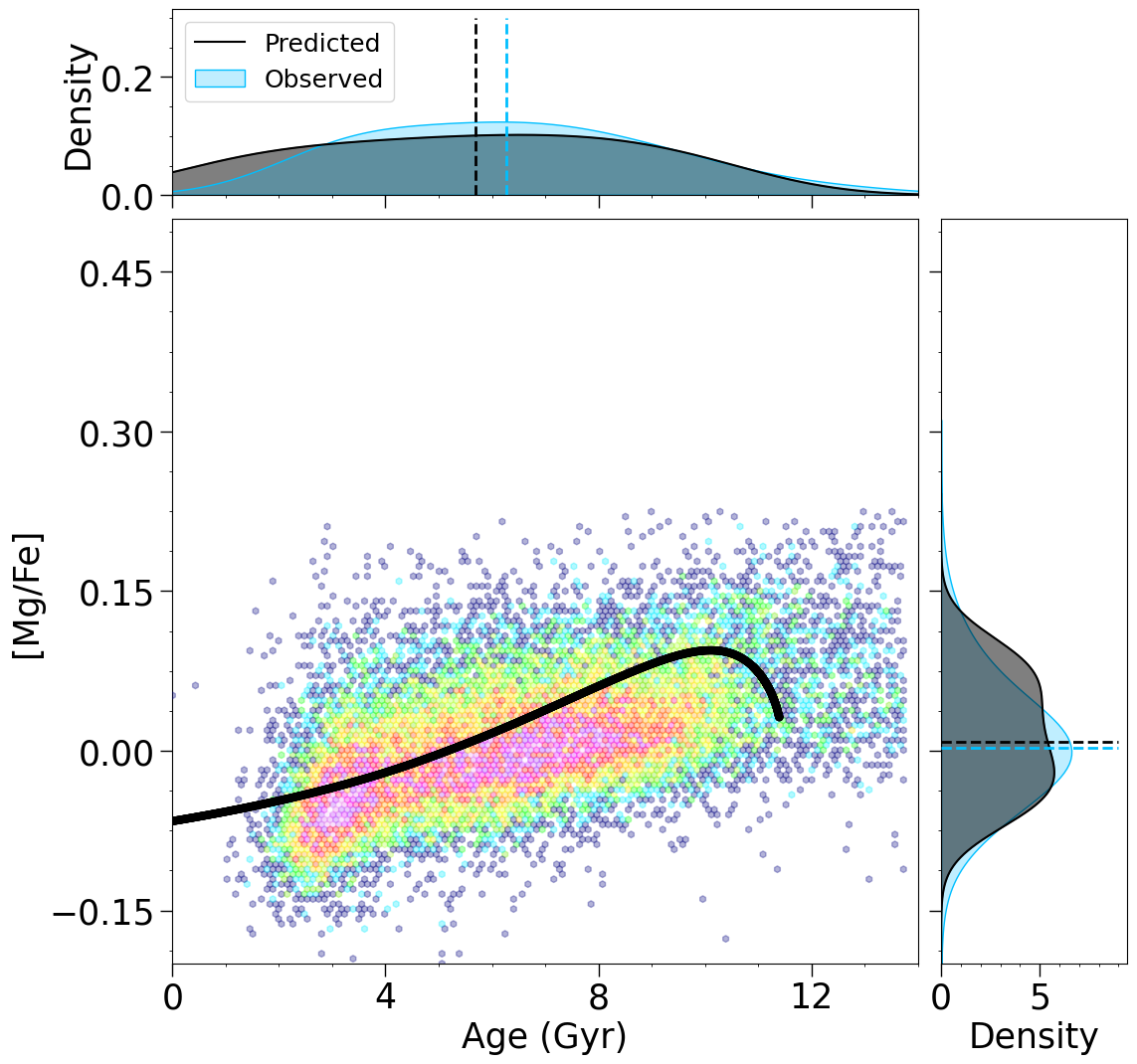} 
                \includegraphics[scale=0.3]{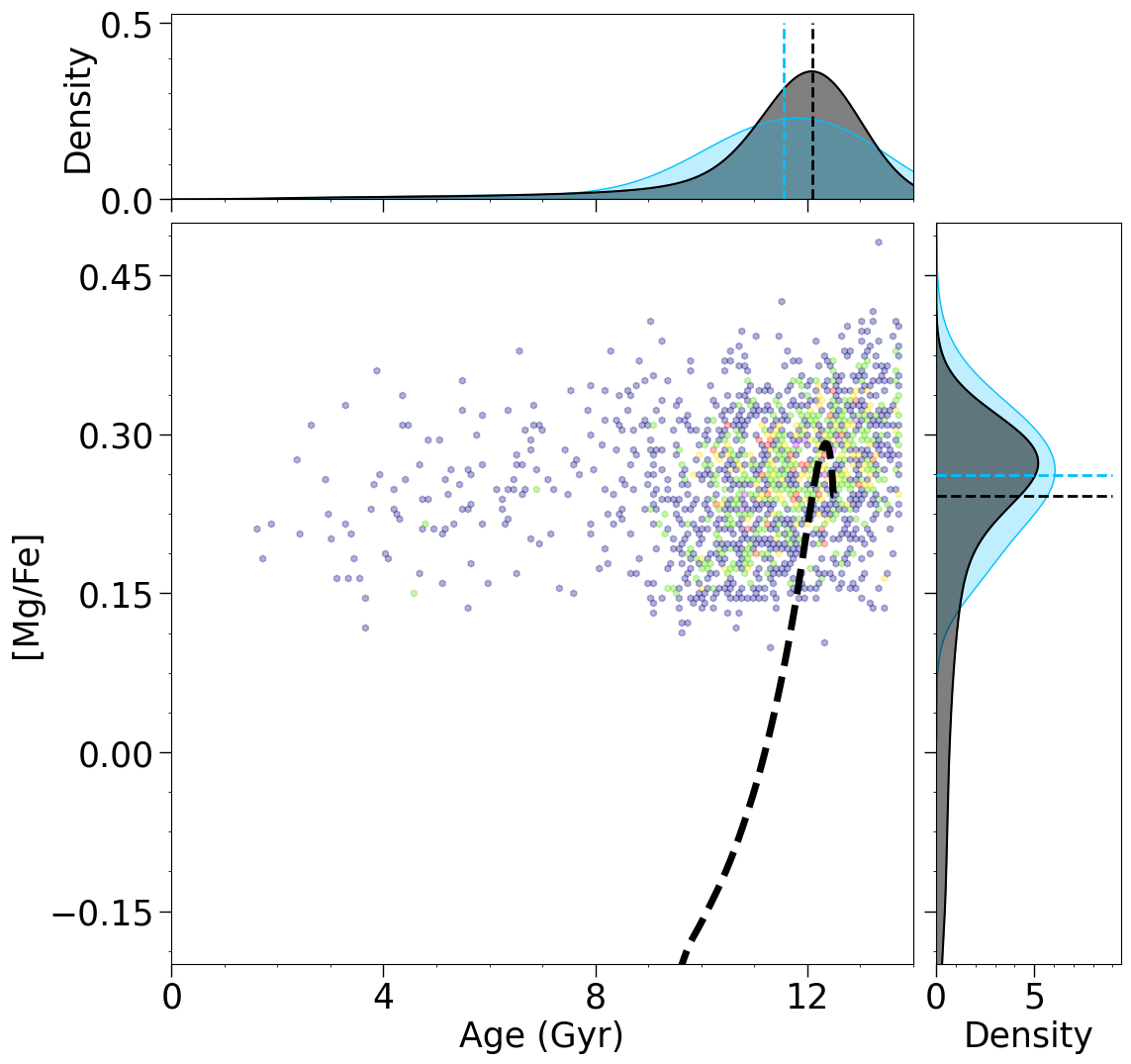} 
    \caption{Observed and predicted [$\alpha$/Fe] vs. age for the low- and high-alpha sequences (left and right panels, respectively). Data are from \cite{Borbolato2025}. The predictions are from the revised parallel model for the Galactic thick and thin discs.
    }
    \label{appendixB2}
\end{figure*}

In Fig. \ref{appendixB}, we show the observed and predicted [$\alpha$/Fe] vs. [Fe/H] and [$\alpha$/Fe] vs. age diagrams for the high- and low-$\alpha$ sequences, separately. In fact, the strenght of the parallel approach is the possibility to follow the evolution of the two components separately. We divide the data from \cite{Borbolato2025} into high- and low-$\alpha$ sequences using a classical chemical separation, as first suggested by \cite{Adibekyan2012}, and we compare to predictions for our chemical evolution models. In the upper panels, we show the observed and predicted [$\alpha$/Fe] vs. [Fe/H] diagrams for the high- (left) and low-$\alpha$ (right) sequences, separately; we can see that the bulk of high- (left) and low-$\alpha$ (right) data are reproduced, as well as their MDFs and [Mg/Fe] distributions, with the thick disc being $\alpha$-enhanced and more metal-poor with respect to the thin disc. To reproduce the metal-rich tail of the thin disc MDF, there can be some radial migration effects and stars coming from the inner part of the Galaxy. With the thick disc model, we have also some thick-disc stars at the higher metallicities, as discussed in \cite{Grisoni2017} even if there are few thick-disc stars at those metallicities; the combined MDF, as shown in Fig. 4, is well reproduced.
\\In Fig. \ref{appendixB2}, we show the observed and predicted [$\alpha$/Fe] vs. age diagrams for the high- (left) and low-$\alpha$ (right) sequences, separately; we can see that the bulk of high- (left) and low-$\alpha$ (right) data are reproduced, as well as their age distributions and [Mg/Fe] distributions, with the thick disc a very old population peaked at 11-12 Gyr ago (as also suggested by several age studies, e.g. \citealt{Miglio2021,Queiroz2023}). The thick disc [Mg/Fe] distribution is reproduced, with few stars reaching also low-[Mg/Fe] values at old ages; those can be the old low-$\alpha$. Alternatively, the old low-$\alpha$ stars can be explained with our chemical evolution model of the thin disc; in fact, the age distribution of the thin disc is well reproduced with our thin disc model.

\end{appendix}

\end{document}